\documentclass[journal]{IEEEtran}

\usepackage{lineno,hyperref}
\usepackage{amsmath}
\usepackage{lineno,multicol}
\usepackage{graphicx}
\usepackage{makecell}
\usepackage{subfig}
\usepackage{lipsum}
\usepackage{float}
\usepackage{epstopdf}
\usepackage{amsthm}
\usepackage{cases}
\usepackage{pbox}
\usepackage{extarrows}
\usepackage{cite}
\usepackage{amssymb}
\usepackage{mathrsfs}

\newtheorem{definition}{Definition}

\if CLASSINFOpdf
\else

\fi

\begin{document}

\title{IF equation: a feature extractor for high-concentration time-frequency representation of mixed signals}
\author{Xiangxiang Zhu,  Kunde Yang,  Zhuosheng Zhang
\thanks{This work was supported by  the Natural Science Foundation of Shaanxi Province under Grant 2023-JC-QN-0739, 
	the Guangdong Basic and Applied Basic Research Foundation under Grant no. 208058579085, the  National Natural Science Foundation of China under Grant no. U20B2075, the Fundamental Research Funds for the Central Universities  under Grant no.  G2021KY05103.}
\thanks{X. Zhu is with School  of Mathematics and Statistics, Northwestern Polytechnical University,  Xi'an 710072, China (e-mail: zhuxiangxiang@nwpu.edu.cn)} 	
\thanks{K. Yang  is with School of Marine Science and Technology, Northwestern Polytechnical University, Xi’an 710072, China (e-mail: ykdzym@nwpu.edu.cn).}
\thanks{ Z. Zhang  is with School  of Mathematics and Statistics, Xi'an Jiaotong University,  Xi'an, 710049, China (e-mail: zszhang@mail.xjtu.edu.cn).}}

\markboth{Time-frequency analysis and its application}%
{Shell \MakeLowercase{\textit{et al.}}: Bare Demo of IEEEtran.cls for IEEE Journals}
\maketitle

\begin{abstract}
High-concentration time-frequency (TF) representation provides a valuable tool for characterizing multi-component non-stationary signals.  In our previous work, we proposed using an instantaneous frequency (IF) equation to sharpen the TF distribution, and experiments verified its effectiveness.   In this paper, we systematically discuss why the IF equation-based TF analysis methods work and how to use the IF equation to improve TF sharpness. By the analysis of the properties of the IF equation, we prove that a good IF equation can unify the well-known  IF and group delay estimators and provides an effective way to characterize the mixture of time-varying and frequency-varying signals. By discussing the post-processing techniques based on the IF equation, we can prove that many popular TF post-processing methods, such as the synchroextracting transform, the multi-synchrosqueezing transform, and the time extracting transform,   fall into the IF equation-based category. We also propose a novel approach to combine different IF equations to minimize energy spreading based on local sparsity.  Numerical simulations and practical experiments are presented to illustrate the performance of the proposed IF equation-based  TF  analysis method. 
\end{abstract}

\begin{IEEEkeywords} High-resolution time-frequency analysis,  Synchrosqueezing transform,   Synchroextracting transform,   IF equation, Mixed signals. 

\end{IEEEkeywords}

\IEEEpeerreviewmaketitle

\section{Introduction}

\IEEEPARstart
The study  and  the interpretation of multi-component non-stationary signals require a powerful  analysis tool.   Time-frequency (TF) analysis methods   are used to measure how a signal's frequency components vary with time and  provide an effective tool  in  non-stationary signals analysis [1-7].  The most frequently used TF representation  is the linear TF analysis method, mainly including the short-time Fourier transform (STFT) [8], the continuous wavelet transform (CWT) [9], the Stockwell transform [10], and the chirplet transform (CT) [11].  This kind of method  characterizes the signal  via inner products with a  pre-assigned basis and  has invertibility and  low computation cost. However, restricted by the Heisenberg uncertainty principle [12],  the linear TF analysis  can not simultaneously obtain optimal time and frequency resolution, easily leading to a blurred TF representation.

To solve the above limitation,  the post-processing methods of the linear TF representations have been widely studied in the past decades.  The reassignment method (RM),  proposed by Kodera et al. to the spectrogram [13]  and generalized by F. Auger and P. Flandrin to the time-scale distribution and others  [14],  reassigns the TF energy from the original position to the gravity center of the signal's energy distribution such that a sharped TF representation is yielded.  F. Auger et al. also extended the standard  RM  to the Levenberg-Marquardt reassignment [15],  making the concentration of the signal energy adjustable.  J. Xiao and  P. Flandrin   [16] presented the multitaper TF reassignment with the two-fold objective of a sharp localization for the chirp components and a reduced level of statistical fluctuations for the noise.  In [17], Bruni et al. proposed a  modified  RM  for analyzing multi-component crossing signals with cross-over instantaneous frequencies (IFs). This method corrects the TF result of the standard  RM  in the non-separable region partly.  In [18],  a  multitaper reassigned spectrogram was proposed for oscillating transients with Gaussian envelopes, which localizes energy at the instantaneous frequency (IF) of transients, while  Gaussian white noise is made flatter by the use of multiple windows. In [19], an iterative reassignment method was presented to extract transient features of strongly time-varying signals and frequency-varying signals. Although  RM  and its variants improve the energy concentration of the linear TF representation,  they lack an explicit formula for signal reconstruction, which limits its application [20].

Another popular TF post-processing method is the synchrosqueezing transform (SST), initialized in  [21]  and further analyzed in  [22].  This method was developed in understanding the principle of the empirical mode decomposition [23]. By squeezing the TF coefficients into the IF trajectory in the frequency direction, SST not only enhances TF readability but also allows for mode retrieval. Because of the advantages of mode representation and reconstruction,   many new developments of SST have been carried out in various directions.  First,  an extension to the STFT case was proposed in [24],  while a generalization of the wavelet approach utilizing wavelet packet decomposition for both one-dimensional and two-dimensional cases was available in [25,26].  Extensions to other transform frameworks included,  but not limited to,  the synchrosqueezing S-transform [27],  the synchrosqueezing three-parameter wavelet transform [28], and the fractional synchrosqueezing transformation [29].  Second, the robust analysis of SST  has been studied in [30-32]  and a new IF estimator within the framework of the signal’s phase derivative and the linear canonical transform was introduced in [33]. Moreover,  a study of SST  applied to multivariate data was done in [34].    An adaptive SST with a time-varying parameter was introduced in [35,36], which obtains the well-separated condition for multi-component signals using the linear frequency modulation to approximate a non-stationary signal at every time instant.  The theoretical analysis of adaptive SST was studied in [37,38].   Finally,  combining the SST  with the ideal TF representation theory,  the synchroextracting transform (SET) [39] was proposed to improve the TF concentration by extracting the ridges of the STFT. 

SST works well for sinusoidal signals but suffers from low TF resolution when dealing with strong modulation signals that consist of a fast-varying feature. To address this drawback,  many improvements have been presented, mainly including the demodulated SST [40,41], the high-order SST  [42,43], the multi-synchrosqueezing transform [44,45], and the time-reassigned SST [46,47].  The demodulated SSTs [40,41]  aim to change broad-band signals to narrow-band signals and the standard  SST  is followed then.  The high-order SSTs  [42,43] aim to calculate an IF estimation in the TF or time-scale domain for more general signal types.  The multi-synchrosqueezing transformations [44,45] apply an iterative reassignment technique to sharpen the blurred TF energy step-by-step. The time-reassigned SSTs [46,47], in fact, introduce a dual operator by considering time reassignment instead of frequency reassignment and have a good performance in addressing pulsive-like signals. A  comprehensive review of the subject of reassigned TF representations is provided in Ref. [5,20].

The mechanism of the SST  and its variants introduced above is to derive an  IF or group delay (GD) estimator for different transform frameworks and different-type signals,  and further to reassign TF coefficients from the original position to the estimated instantaneous trajectories. Indeed, the IF (or GD) estimator is critical to this kind of method (we can denote it as the IF estimator-based TF post-processing method) because it determines the accuracy of high-concentration TF representations. Instead of computing the IF (or GD) estimation,  another high-concentration TF representation is based on the  IF equation [48-50], i.e.,  a non-linear function in the TF domain with the IFs of the signal as its solution. This kind of method (denoted as the IF equation-based TF post-processing method) enhances the TF concentration by solving the IF equation using the fixed point iterative algorithm [48] or the extractor [49,50] and has been used in the sound signal analysis [48,49], mechanical signal processing [51], and medical data analysis [52].  Several issues about the IF equation-based TF post-processing methods, however,  should be considered.  First,   theoretical analysis of the IF equation and how to use it to concentrate the TF distribution should be explained more comprehensively, involving the definition and properties of the IF equation, and the convergence of its solution algorithm. Second, we know an IF estimator or GD estimator can not simultaneously characterize the time-varying and frequency-varying information of a mixed signal. How about the IF equation?  Last but not least, we should consider if there is a  theoretical model to cover the prevailing TF analysis methods, including the SET [39], the multi-synchrosqueezing transformation [44],  and the time synchroextracting transform [52].

In this paper, we focus on the  IF equation-based  TF post-processing methods and attempt to build a theoretical framework for it.  We first review the IF estimator-based TF post-processing methods and then define the IF equation followed by three concrete expressions under the STFT framework. We study the properties of the IF equation,  and introduce two ways, that is, by the extractor to detect signal information and by iterative algorithms for TF reassignment,   to concentrate the TF localization using the IF equation.  The difference between the IF estimator-based methods and the  IF equation-based methods is discussed. By analysis, a significant advantage of the IF equation-based methods over the former is that a good IF equation contains not only the IF information but also the GD information, which can unify the IF  and GD  estimators and characterizes mixed signals effectively.  By combining two IF equations,  a novel  TF analysis approach is proposed to further minimize energy spread based on local sparsity.  Finally,  numerical examples of the theoretical derivations are presented. The potential applications are also explored and verified.   

The remainder of the paper is organized as follows. In Section \textrm{II}, we review the IF estimator-based TF post-processing methods.    Section  \textrm{III}  presents the proposed IF equation-based  TF post-processing methods with definition, properties, and post-processing technologies.  Section  \textrm{IV} proposes a combination of extracting transform with different IF equations.   Comparative studies and applications are presented in Section \textrm{V}. Finally, the conclusions are drawn in Section \textrm{VI}.

\section{A review of the IF estimator-based TF post-processing methods}  

\subsection{Signal model}
Multi-component non-stationary signals are comprised of several mono-components and form the general case. In this paper we use the  popular amplitude-modulation and frequency-modulation  (AM-FM) model  to  describe   the time-varying  features of multi-component non-stationary signals, which is defined as 
\begin{equation}
	f(t)=\sum_{k=1}^{K}f_k(t)=\sum_{k=1}^{K}A_k(t)\mathrm{e}^{j\phi_k(t)}, 
\end{equation}
where $K$ is a positive integer representing the number of AM-FM components, $j$ denotes the imaginary unit  with   $j^2=-1$,  $A_k(t)>0$  and $\phi_k(t)$   denote   the instantaneous amplitude and  instantaneous phase of the $k$-th component (or mode), respectively. The first and the second  derivatives of the  phase,  i.e.,   $\phi'_k(t)$, $\phi''_k(t)$, are referred to as  the instantaneous  frequency (IF) and chirp rate (CR) of the $k$-th component.


\subsection{IF estimator-based TF post-processing method}
The TF post-processing methods, originating from the RM [14] and SST [22],  relocate the original TF points to concentrate the spread  TF distribution. With this regard, a crucial problem is to determine where the blurred  TF  points are rearranged. The ideal one,  of course,  is moved to the true  IF curves of the signal analyzed, which thus drives the study  of the IF estimators in the TF domain. Under the  STFT framework, we give an introduction to the IF estimator-based TF post-processing methods.  

Given signal  $f(t)\in L^2(\mathbb{R})$, its Fourier transform is defined as
\begin{equation}
	\hat{f}(\omega)=\int_{-\infty}^{+\infty}f(t)\mathrm{e}^{-j\omega t} \mathrm{d}t.
\end{equation}
The  (modified) STFT  of $f(t)$,  which is obtained through the use of a sliding window $g(t)\in L^2(\mathbb{R})$,  is given by 
\begin{equation}
	\begin{split}
		S_{f}^{g}(t, \omega)=\int_{-\infty}^{+\infty}f(\mu)g^*(\mu-t) \mathrm{e}^{-j\omega(\mu- t)}\mathrm{d}\mu,
	\end{split}
\end{equation}
 where the superscript $*$ denotes the complex conjugate.  It is known that the STFT can not simultaneously localize a signal in both time and frequency precisely due to the limitation of the  Heisenberg uncertainty principle [12], i.e., a long window favors better concentration in frequency but worse temporal localization, while a short window provides good temporal localization but poor frequency resolution. To improve the TF localization precision and show good TF readability,  the  TF post-processing methods [22,24,25,27,42,43] are presented to relocate the TF coefficients into the IF trajectories.  Thanks to  G. Thakur and  H.T. Wu, they develop  I. Daubechies's work [22] and give a two-dimensional  IF estimator by STFT   as follows [24]:
\begin{equation}
	\hat{\omega}_{1}(t,\omega)=\Re\left\lbrace \frac{\frac{\partial}{\partial t}S^g_f(t,\omega)}{jS^g_f(t,\omega)}\right\rbrace,
\end{equation}
here $|S^g_f(t,\omega)|>\lambda$ ($\lambda>0$),  and  $\Re\{\cdot\}$ is the real  part of complex number. The IF estimator  (4)  is accurate for the harmonic-like  signals with the stationary  instantaneous  feature,  i.e., each model of the  signal $f(t)$ satisfies $|A'_k(t)|<\epsilon$, $|\phi''_k(t)|<\epsilon$ for $k=1,2,\cdots,K$ (here $\epsilon$ is a small positive constant).

 With the  IF estimator $\hat{\omega}_{1}(t,\omega)$ defined by (4), the STFT-based SST (FSST) [24] is to reassign the TF coefficients from the original location  to the estimated IF trajectory  along  the frequency direction:  
\begin{equation}
	T_f^1(t,\omega)=\int_{|S^g_f(t,\eta)|>\lambda}S^g_f(t,\eta)\delta (\omega-\hat{\omega}_{1}(t,\eta))\mathrm{d}\eta, 
\end{equation}
where  $\delta(\cdot)$  is  the  Dirac delta function.

FSST suffers from a low TF resolution at the modulated part of fast-varying signals, such as chirp-like signals.  To  address  this  shortcoming, paper [42] introduced  another IF estimator  as (6), where $\tilde{q}(t, \omega)$ is given by  
\begin{figure*}[bhpt] 
\begin{align}
		\hat{\omega}_{2}(t,\omega)=
		\begin{cases}
			\Re\left\lbrace \frac{\frac{\partial}{\partial t}S^g_f(t,\omega)}{jS^g_f(t,\omega)}\right\rbrace+\Re\left\lbrace \tilde{q}(t, \omega) \frac{\frac{\partial}{\partial \omega}S^g_f(t,\omega)}{S^g_f(t,\omega)}\right\rbrace,   & ~\text{if}~ \frac{\partial}{\partial t} \left( \frac{\frac{\partial }{\partial \omega} S^g_f(t,\omega)}{S^g_f(t,\omega)} \right) \neq j,   \text{and}~ S^g_f(t,\omega)\neq 0, \\
			\hat{\omega}_{1}(t,\omega),  & ~\text{if}~ \frac{\partial}{\partial t} \left( \frac{\frac{\partial }{\partial \omega} S^g_f(t,\omega)}{S^g_f(t,\omega)} \right) = j,   \text{and}~ S^g_f(t,\omega)\neq 0, 
		\end{cases}
\end{align}
\end{figure*}	
\begin{equation}
	\tilde{q}(t, \omega)=\frac{\frac{\partial}{\partial t} \left( \frac{\frac{\partial }{\partial t} S^g_f(t,\omega)}{S^g_f(t,\omega)} \right)}{j-\frac{\partial}{\partial t} \left( \frac{\frac{\partial }{\partial \omega} S^g_f(t,\omega)}{S^g_f(t,\omega)} \right)}.
\end{equation}
The quantity $\hat{\omega}_{2}(t,\omega)$ corresponds to the true IFs  of the Gaussian modulated  linear chirp [42]. The second-order SST then consists in replacing $\hat{\omega}_{1}(t,\omega)$ by $\hat{\omega}_{2}(t,\omega)$ in the standard FSST:
\begin{equation}
	T_f^2(t,\omega)=\int_{|S^g_f(t,\eta)|>\lambda}S^g_f(t,\eta)\delta (\omega-\hat{\omega}_{2}(t,\eta))\mathrm{d}\eta.
\end{equation}

In fact,  the  IF estimator-based TF post-processing methods  can  be unified  as 
\begin{equation}
	T_f(t,\omega)=\int_{|S(t,\eta)|>\lambda}S(t,\eta)\delta (\omega-\hat{\omega}(t,\eta))\mathrm{d}\eta, 
\end{equation}
where   $S(t,\eta)$ denotes a  TF transform, and $\hat{\omega}(t,\eta)$ is an  IF estimator calculated  in  the TF domain. 
If the GD estimation $\hat{t}(\mu,\omega)$ under the  TF representation  is known, the   reassignment of  the TF coefficients  along  the time direction can be  similarly performed: 
\begin{equation}
	\tilde{T}_f(t,\omega)=\int_{|S(\mu,\omega)|>\lambda}S(\mu,\omega)\delta (t-\hat{t}(\mu,\omega))\mathrm{d}\mu.
\end{equation}

The IF estimator-based TF methods are all that an IF (or GD) estimator is determined first in the transform domain and then a relocation step is performed.   Different from the mechanism of this kind of method, in the following, we will introduce another class method, i.e., the IF equation-based TF post-processing method. This method can provide a high-concentration TF representation for the mixed signals.

\section{IF equation-based TF post-processing method}

The central idea of the IF equation-based TF post-processing method is to find an IF equation in the TF domain to characterize the instantaneous features of the signal. By solving the  IF  equation,  it is expected to output a concentrated and accurate TF representation.

\subsection{Definition  of IF equation}
\begin{definition}
	For a multi-component non-stationary signal  $f(t)=\sum_{k=1}^{K}A_k(t)\mathrm{e}^{j\phi_k(t)}$,  let $\Omega$ be a set that  $\Omega\subset\{(t,\omega)|t\in \mathbb{R}, \omega >0\}$  and  $(t,\phi'_k(t))\in \Omega$,  $t\in \mathbb{R}$, $k=1,2,\cdots, K$.  A binary  equation $h_f(t,\omega)=0$ is called the  IF equation of  $f(t)$ on $\Omega$ if, for any  $(t, \omega)\in \Omega$,  it  satisfies  
\begin{equation}
	\begin{split}
		&h_f(t,\omega)=0 ~\text{if } ~\omega=\phi'_k(t) ~\text{for an integer $k \in \{1,2,\cdots, K\}$}, \\
		&  h_f(t,\omega)\neq 0  ~\text{if} ~\omega\neq\phi'_k(t) ~\text{for any integer $k \in \{1,2,\cdots, K\}$}.
	\end{split}
\end{equation}
\end{definition}
From Definition 1, we know that the IF curves $\omega=\phi'_k(t)$ of each model of $f(t)$ are the solutions of its IF equation. Therefore,  the IF equation, including the main information of the studied signal, can be considered a feature extractor.   

In order to show the IF equation more clearly, we now consider a specific example of constant amplitude and linear frequency modulation signal (also known as chirp), i.e., $f(t)=Ae^{j\phi (t)}$,  where $A$ is a positive constant,  $\phi(t)=a+bt+\frac{c}{2}t^2$,  $a, b, c$ are real numbers. In this case, the STFT under the Gaussian window  $g(t)=\frac{1}{\sqrt{2\pi}\sigma}e^{-\frac{t^2}{2\sigma^2}}$  is calculated as 
\begin{align}
	& S_{f}^{g}(t, \omega)=\int_{-\infty}^{+\infty}f(\mu)g^*(\mu-t) \mathrm{e}^{-j\omega(\mu- t)}\mathrm{d}\mu \nonumber\\& =f(t) \frac{1}{\sqrt {1-j\sigma^2c}}\exp\left( -\frac{\sigma^2}{2}\frac{(\omega-\phi'(t))^2}{1-j\sigma^2 c} \right),
\end{align}
where $\phi'(t)=b+ct$. From (12),  we can obtain that
\begin{equation}
	\begin{split}
		\frac{\partial}{\partial \omega}S_{f}^g(t, \omega)=S_{f}^g(t, \omega)\left( -\frac{\sigma^2(\omega-\phi'(t))}{1-j\sigma^2c} \right).
	\end{split}
\end{equation}
The set of points $\omega=\phi'(t)$ satisfies $ \left| \frac{\frac{\partial}{\partial \omega}S_{f}^g(t, \omega)}{S_{f}^g(t, \omega)} \right|^2 =0 \text{~~for} ~| S_{f}^g(t, \omega)|>\lambda$ (here  $0<\lambda <\frac{A}{2\sqrt{1+\sigma^4c^2}}$).
Therefore, we can define an  IF equation  of  $f(t)=Ae^{j\phi (t)}$  on  $\Omega=\{(t,\omega)|t\in \mathbb{R}, \omega >0, | S_{f}^g(t, \omega)|>\lambda\}$ as 
\begin{equation}
IF_1:  ~~ \left| \frac{\frac{\partial}{\partial \omega}S_{f}^g(t, \omega)}{S_{f}^g(t, \omega)} \right|^2=0.
\end{equation}

From (13), it is easy to see that  the  other  two  IF equations  of the chirp signal $f(t)$ on $\Omega$  are  formulated as 
\begin{gather}
IF_2:  ~~	
	\Re\left\lbrace  \frac{\frac{\partial}{\partial \omega}S_{f}^g(t, \omega)}{S_{f}^g(t, \omega)}\right\rbrace=0,\\
IF_3:  ~~	 
	\Im\left\lbrace  \frac{\frac{\partial}{\partial \omega}S_{f}^g(t, \omega)}{S_{f}^g(t, \omega)}\right\rbrace=0, 
\end{gather}
where $\Im\{\cdot\}$ is the imaginary  part of complex number. Therefore, the IF equation of the same signal is not unique. Various IF equations have different properties. By the numerical experiments (see Fig. 1), we will see that $IF_2$ is more appropriate to characterize the time-varying signals, whereas $IF_3$  is more effective when handling the frequency-varying signals.  Thus,  finding a powerful  IF equation that covers as many signal types as possible and has some good properties for its solutions is an important work that needs attention and exploration.    

 The  three equations can be more efficiently computed by
\begin{equation}
	\begin{split}
	\frac{\partial}{\partial \omega}S_{f}^g(t, \omega)=S_{f}^{tg}(t, \omega) \text{~~for} ~(t, \omega)\in \Omega, 
	\end{split}
\end{equation}
where $S_{f}^{tg}(t, \omega)$ denotes the STFT of $f(t)$ using $tg(t)$  as its window. It is also noted that other TF transformations or  window functions can  be used to define the IF equation. Indeed,   we can define the IF equation of chirp signal  similar to  $IF_1$, ($IF_2$ or  $IF_3$)  if an even  window function  is employed.  

In the following section, using  $IF_1$  as the main research object,  we discuss some properties of the  IF equation.  
\subsection{Properties of the  IF equation}
1) \textit{Existence and uniqueness:}  The definition of the IF equation indicates that, for each time $t$, $\omega=\phi'(t)$ is the solution of  $IF_1$  over the set  $\Omega$. If considering a mono-component chirp,  $\omega=\phi'(t)$ is the unique solution for each time $t$.  This property is essential for detecting the signal feature because that can characterize the instantaneous information of the signal and exclude other unwanted information.   

2) \textit{Impulse signals detection:} The instantaneous information $t=t_0$ of impulse signal $f(t)=A\delta(t-t_0)$ is the  unique  solution of $IF_1$ for the fixed $\omega$.

The STFT of the transient signal $f(t)=A\delta(t-t_0)$  with the Gaussian window  is  computed as 
\begin{align} 
	S_{f}^{g} (t,\omega) =A{(\sqrt{2\pi}\sigma)}^{-1}\mathrm{e}^{-\frac{(t_0-t)^2}{2\sigma^2}}\mathrm{e}^{-j\omega(t_0-t)}.
\end{align}        
From (18), we can obtain that 
\begin{equation}
	\begin{split}
		\left| \frac{\frac{\partial}{\partial \omega}S_{f}^g(t, \omega)}{S_{f}^g(t, \omega)}\right|^2 =(t-t_0)^2.
	\end{split}
\end{equation}	
Therefore, for any $\omega$, $t=t_0$  is the unique solution of   $IF_1$.  

The $IF_3$ also can characterize this impulse signal,  whereas  $IF_2$ does not include the transient feature of the signal.  Consequently, the  $IF_1$ and    $IF_3$   are more preferred than  $IF_2$ when dealing with impulse signals.  

3) \textit{Detection of transient signal with Gaussian envelope:} The TF centre ($t_0$, $\omega_0$) of an oscillating transient with a Gaussian envelope $f(t)=A\mathrm{e}^{-\frac{(t-t_0)^2}{2\sigma_0^2}}\mathrm{e}^{j\omega_0t}$ is the unique solution of $IF_1$.

The STFT of  signal  $f(t)=A\mathrm{e}^{-\frac{(t-t_0)^2}{2\sigma_0^2}}\mathrm{e}^{j\omega_0t}$   is  computed as 
\begin{equation}
	\begin{split}
	&	S_{f}^{g}(t,\omega)\\&={(\sqrt{2\pi}\sigma)}^{-1}{f}(t)\int_{-\infty}^{+\infty}\mathrm{e}^{-\frac{(t-t_0)\mu}{\sigma_0^2}}\mathrm{e}^{-\frac{\mu^2}{2}(\frac{1}{\sigma_0^2}+\frac{1}{\sigma^2})} \mathrm{e}^{-j(\omega-\omega_0)\mu}\mathrm{d}\mu,\\
		&={f}(t)\mathrm{e}^{\frac{\sigma^2}{2\sigma_0^2(\sigma^2+\sigma_0^2)}(t-t_0)^2}\mathrm{e}^{-\frac{\sigma^2\sigma_0^2}{2(\sigma^2+\sigma_0^2)}(\omega-\omega_0)^2}\mathrm{e}^{j\frac{\sigma^2}{\sigma^2+\sigma_0^2}(t-t_0)(\omega-\omega_0)}.
	\end{split}
\end{equation}
From (20), it follows that 
\begin{align} 
	\left| \frac{\frac{\partial}{\partial \omega}S_{f}^g(t, \omega)}{S_{f}^g(t, \omega)}\right|^2= \left|-\frac{\sigma^2\sigma_0^2}{\sigma^2+\sigma_0^2}(\omega-\omega_0) + j\frac{\sigma^2}{\sigma^2+\sigma_0^2}(t-t_0)\right|^2, 
\end{align}       
Thus the  point  ($t_0$, $\omega_0$) is the unique solution of  $IF_1$  to make $h_f(t, \omega):=	\left| \frac{\frac{\partial}{\partial \omega}S_{f}^g(t, \omega)}{S_{f}^g(t, \omega)}\right|^2=0$.

4) \textit{Linear group delay detection:} For any $\omega$, the  GD of linear group delay  signal $\hat{f}(\omega)=Ae^{-j(a+b\omega+\frac{c}{2}\omega^2)}$ is the   unique  solution of  $IF_1$. 

Since the STFT of $f(t)$ can be rewritten as  
\begin{equation}
	\begin{split}
		S_{f}^{g}(t, \omega)&=\int_{-\infty}^{+\infty}f(\mu)g^*(\mu-t) \mathrm{e}^{-j\omega(\mu- t)}\mathrm{d}\mu\\
		&=\frac{1}{2\pi}\int_{-\infty}^{+\infty}\hat{f}(\nu)\hat{g}^*(\nu-\omega) \mathrm{e}^{jt\nu}\mathrm{d}\nu, 
	\end{split}
\end{equation}
then the STFT of linear group delay  model  can be expressed as 
\begin{equation}
	\begin{split}
		S_{f}^{g}(t,\omega)=\frac{1}{2\pi}\hat{f}(\omega)\mathrm{e}^{jt\omega}\int_{-\infty}^{+\infty}\mathrm{e}^{-j\frac{c}{2}\nu^2}\hat{g}^*(\nu) \mathrm{e}^{j(t-b-c\omega)\nu}\mathrm{d}\nu.
	\end{split}
\end{equation}
From (23), we have 
\begin{equation}
	\begin{split}
		&\frac{\partial}{\partial \omega}S_{f}^g(t, \omega)=j(t-b-c\omega)S_{f}^g(t, \omega)\\&-\frac{jc}{2\pi}\hat{f}(\omega)\mathrm{e}^{jt\omega}\int_{-\infty}^{+\infty}\mathrm{e}^{-j\frac{c}{2}\nu^2}\nu \hat{g}^*(\nu) \mathrm{e}^{j(t-b-c\omega)\nu}\mathrm{d}\nu.
	\end{split}
\end{equation}
Due to the  Fourier transform of Gaussian window $g(t)=\frac{1}{\sqrt{2\pi}\sigma}e^{-\frac{t^2}{2\sigma^2}}$ is calculated as  
\begin{equation}
	\hat{g}(\nu)=\mathrm{e}^{-\frac{\sigma^2}{2}\nu^2}, 
\end{equation}
which yields that 
$$\int_{-\infty}^{+\infty}\mathrm{e}^{-j\frac{c}{2}\nu^2}\nu \hat{g}^*(\nu) \mathrm{d}\nu=0.$$
Consequently,  $t=b+c\omega$ satisfies $h_f(b+c\omega,\omega)=0$. The uniqueness of the solution  can  be  proved  by deducing the expression  of (23)  using  (25) for the linear group delay  signal (one can refer to the calculation of (12)). 

This result shows a clear advantage of the IF equation over the IF estimator since  $IF_1$ covers the instantaneous characteristic of both the chirp signal and the linear group delay model, unifying the well-known  IF and GD  estimators proposed in the SST [22,24] and the time-reassigned SST [46].

5) \textit{Characterizing a mixture of harmonic-like and pulsive-like signals:} Let      $f(t)=A_1\mathrm{e}^{j\omega_1t}+A_2\mathrm{e}^{-\frac{(t-t_2)^2}{2\tilde{\sigma}^2}}\mathrm{e}^{j\omega_2t}$ be a multi-component signal consisting of harmonic  and pulse  components.  Suppose  the support of  $\mathrm{e}^{-\frac{\tilde{\sigma}^2}{2}\omega^2}$ be $[-\Delta, \Delta]$, and the parameter of $\sigma $ of the Gaussian window used in the STFT be taken as ${\sigma}=l\tilde{\sigma}$, here $l>0$. If $\left| \omega_{1}-\omega_{2}\right|>\frac{\sqrt{1+l^2}}{l}\Delta$,  then 
$$h_f(t, \omega_1)=0 ~  \text{and} ~  h_f(t_2, \omega_2)=0.$$

The STFT of the two-component signal $f(t)$ under the Gaussian window  is  that  
\begin{equation}
	\begin{split}
	&	S_{f}^{g}(t,\omega)=f_1(t)\mathrm{e}^{-\frac{l^2\tilde{\sigma}^2}{2}(\omega-\omega_1)^2}+\frac{A_2}{\sqrt{1+l^2}}\mathrm{e}^{-\frac{l^2\tilde{\sigma}^2}{2(1+l^2)}(\omega-\omega_2)^2}\times\\&\mathrm{e}^{-\frac{(t-t_2)^2}{2(1+l^2)\tilde{\sigma}^2}}\mathrm{e}^{j\frac{l^2}{1+l^2}(\omega-\omega_2)(t-t_2)}\mathrm{e}^{j\omega_{2}t}.
	\end{split}
\end{equation}
Thus, we have  
\begin{equation}
	\begin{split}
		&\frac{\partial}{\partial\omega}S_{f}^g(t,\omega)=-l^2\tilde{\sigma}^2f_1(t)(\omega-\omega_1)\mathrm{e}^{-\frac{l^2\tilde{\sigma}^2}{2}(\omega-\omega_1)^2}\\&-\frac{A_2l^2\tilde{\sigma}^2}{\sqrt{1+l^2}(1+l^2)}(\omega-\omega_2)\mathrm{e}^{-\frac{l^2\tilde{\sigma}^2}{2(1+l^2)}(\omega-\omega_2)^2}\mathrm{e}^{-\frac{(t-t_2)^2}{2(1+l^2)\tilde{\sigma}^2}}\times\\&\mathrm{e}^{j\frac{l^2}{1+l^2}(\omega-\omega_2)(t-t_2)}\mathrm{e}^{j\omega_{2}t}+\frac{jA_2l^2}{\sqrt{1+l^2}(1+l^2)}\mathrm{e}^{-\frac{l^2\tilde{\sigma}^2}{2(1+l^2)}(\omega-\omega_2)^2}\times\\&(t-t_2)\mathrm{e}^{-\frac{(t-t_2)^2}{2(1+l^2)\tilde{\sigma}^2}}\mathrm{e}^{j\frac{l^2}{1+l^2}(\omega-\omega_2)(t-t_2)}\mathrm{e}^{j\omega_{2}t}.
	\end{split}
\end{equation}
Due to  
\begin{equation}
	\begin{split}
		&\left| \frac{\partial}{\partial\omega}S_{f}^g(t,\omega_1)\right| 	\leq\frac{A_2l^2\tilde{\sigma}^2}{(1+l^2)^{\frac{3}{2}}}|\omega_1-\omega_2|\mathrm{e}^{-\frac{l^2\tilde{\sigma}^2}{2(1+l^2)}(\omega_1-\omega_2)^2}\\&+\frac{A_2l^2}{(1+l^2)^{\frac{3}{2}}}\mathrm{e}^{-\frac{l^2\tilde{\sigma}^2}{2(1+l^2)}(\omega_1-\omega_2)^2}|t-t_2|\mathrm{e}^{-\frac{(t-t_2)^2}{2(1+l^2)\tilde{\sigma}^2}}\nonumber\\
		&\leq C\mathrm{e}^{-\frac{l^2\tilde{\sigma}^2}{2(1+l^2)}(\omega_1-\omega_2)^2}, 
	\end{split}
\end{equation}
where $C=\frac{A_2l^2\tilde{\sigma}^2}{(1+l^2)^{\frac{3}{2}}}|\omega_1-\omega_2|+\frac{A_2l^2\tilde{\sigma}}{\mathrm{e}^{\frac{1}{2}}(1+l^2)}$. The fact that  $\left| \omega_{1}-\omega_{2}\right|>\frac{\sqrt{1+l^2}}{l}\Delta$ and the support of  $\mathrm{e}^{-\frac{\tilde{\sigma}^2}{2}\omega^2}$ is $[-\Delta, \Delta]$ leads to $\left| \frac{\partial}{\partial\omega}S_{f}^g(t,\omega_1)\right| =0$, therefore we have  $h_f(t, \omega_1)=0$.

Similarly, it can be obtained that 
\begin{equation}
	\begin{split}
		\left| \frac{\partial}{\partial\omega}S_{f}^g(t_2,\omega_2)\right| =l^2\tilde{\sigma}^2A_1|\omega_2-\omega_1|\mathrm{e}^{-\frac{l^2\tilde{\sigma}^2}{2}(\omega_2-\omega_1)^2},	\nonumber
	\end{split}
\end{equation}
which also yields  $h_f(t_2, \omega_2)=0$. Therefore, $IF_1$  extracts the instantaneous information of signals consisting of a mixture of harmonic-like and pulsive-like components.

6) \textit{Combination of IF equations:}  It is known that a single window is difficult to balance time resolution and frequency resolution. The combination of multiple spectrograms or wavelet transforms with short and long windows or a set of wavelets has been shown to result in good joint TF resolution for signals consisting of mixtures of tones and pulses [50,54]. Inspired by this idea, we  can  also define  a multi-resolution IF equation of chirp (impulse or linear group delay) signal:
\begin{equation}
	\begin{split}
		MrIF(t, \omega):=\left|\left( \prod_{i=1}^m \frac{\frac{\partial}{\partial \omega}S_f^{g_{i}}(t, \omega)}{S_{f}^{g_{i}}(t, \omega)}\right)  ^{\frac{1}{m}}\right| =0, 
	\end{split}
\end{equation}
for $| S_{f}^{g_{i}}(t, \omega)|>\lambda$. Here  $S_f^{g_{i}}(t, \omega)$ denotes the STFT of $f(t)$ using  the Gaussian window with parameter $\sigma_i$. This equation corresponds to the geometric mean of multiple IF equations and is built on high-resolution TF  representation. In this regard,  it can be used to characterize the signal with closely-spaced instantaneous information [50]. It should be pointed out that,   except for geometric mean,  one can explore other combination forms of  IF equations for non-stationary signals analysis. 

\subsection{TF post-processing method using  IF equation}

This section  introduces how to use the given IF equation $h_f(t,\omega)=0$ to concentrate the spread TF distribution. 
Generally, there are two classical ways to achieve such goal: extraction  and reassignment. In the following we will discuss these two ways  in detail.

\subsubsection{Extracting  transform}   

The main idea of extracting transform is to present the TF coefficients consistent with the solutions of the IF equation, making the TF location concentrated and matching the IF or GD estimates.

Define an extractor  as
\begin{equation}
	\begin{split}
		\delta(\omega)=
		\begin{cases}
			1,   & ~\text{if}~ \omega= 0, \\
			0,  & ~\text{if}~ \omega\neq 0,
		\end{cases}
	\end{split}
\end{equation}
then  a concentrated TF representation is obtained  by  keeping only the TF points satisfying  the  IF equation,   which is  given by
\begin{equation}
	\begin{split}
		T_e(t, \omega)=S_{f}^g(t, \omega)	\delta(h_f(t,\omega)).
	\end{split}
\end{equation}
We refer to (30) as the extracting transform. This transform presents the oscillation content of the studied signal by detecting the solutions of the IF equation,  retaining a small part of the TF coefficients of STFT and thus greatly sharpening the energy distribution.  

The popular synchroextracting transform (SET) [39] can be considered a specialization of the extracting transform. Indeed, from expression (12), it is easy to get:
\begin{equation}
	\hat{\omega}_1(t,\omega)=\Re \left\lbrace \frac{\frac{\partial}{\partial t}S^g_f(t,\omega)}{jS^g_f(t,\omega)}\right\rbrace =\phi'(t)+\frac{\sigma^4{c}^2(\omega-\phi'(t))}{1+\sigma^4{c}^2}.
\end{equation}
Expression  (31) indicates  that the set of points  $\omega=\phi'(t)$ satisfies
\begin{equation}
	\omega-\hat{\omega}_1(t,\omega)=0.
\end{equation}
Thus, equation (32)  is a specific  IF equation of chirp for $(t, \omega)\in \Omega$.  Based on equation  (32), it is easy to define the  extracting transform  as 
\begin{equation}
	\begin{split}
		T_e(t, \omega)=S_{f}^g(t, \omega)	\delta(\omega-\hat{\omega}_1(t,\omega)),
	\end{split}
\end{equation}
which corresponds to the SET given in  [39]. 

It is known that the extracting transform is not a type of energy conservation because they only retain the signal energy on the TF curves satisfying the IF equation, which brings a challenge to reconstruct the original signal.  Based on equation (12) and  the fact that 
\begin{equation}
	\begin{split}
		T_e(t, {\phi}'(t))=S_{f}^g(t, {\phi}'(t)), 
	\end{split}
\end{equation}
where ${\phi}'(t)$ is the IF and generally estimated by the ridge of the extracting transform in terms of the locally maximal curve of $|T_e(t, \omega)|$,  one can recover the harmonic-like signals from the extracting transform accurately.  However,  the reconstruction performance degrades when recovering the fast-varying signals [49].  Many improvements, such as the SECT [49] and adaptive signal separation operation (ASSO) [55], are proposed to derive a more accurate component recovery formula. 

Another important problem for extracting transform is its numerical implementation,  which should robustly remove the non-information TF points concerning the TF energy diffusion and the noise. The numerical implementation of the extracting transform  is  frequently expressed  as
\begin{equation}
	\begin{split}
		T_e[m, n]= \begin{cases}
			S_{f}^g[m, n],   & \text{$ \left|h_f[m, n]\right|<\frac{\Delta\omega}{2}$}, |S_{f}^g[m, n]|>\lambda,\\
			0,  & \text{otherwise,}
		\end{cases}
	\end{split}
\end{equation}
where  $S_{f}^g[m, n]$ and $h_f[m, n]$ denote the discrete  STFT and IF equation of the sampled signal respectively, and  $\Delta\omega$ is the discrete frequency interval. Except for this way, it is interesting to consider the nonlinear transformation of $h_f$, such as $\exp(h_f)-1$, to obtain a more robust implementation.

\subsubsection{TF reassignment based on   iterative  algorithm}

Another way to sharpen the spread TF distribution is to move the TF values towards the IF or GD curves. Since the IFs or  GDs of the studied signal $f(t)$ are the solutions of the given IF equation $h_f(t,\omega)=0$, i.e., the roots of $h_f(t,\omega)$, then for any  ($t, \omega$) $\in \Omega$,  we can consider it as an initial point of a numerical algorithm to solve a root of $h_f(t,\omega)$.   When the numerical algorithm converges to the solutions of the IF equation with increasing iteration times, the spreading TF energy can be relocated to the instantaneous curves of the studied signal.

There are in fact many algorithms, such as the fixed point algorithm, the Newton algorithm, and the Levenberg-Marquardt (LM) algorithm [55] can be used to obtain a solution of that 
\begin{align}
	h_f(t, \omega)=0. 
\end{align}
Specifically, starting  with an arbitrary initial point ($t, \omega_0$) $\in \Omega$,  the fixed point algorithm yields the  iteration with respect to variable $\omega$ that  
\begin{align}
	\omega_{k+1}=\omega_{k}+ \tau h_f(t, \omega_{k}),  
\end{align}
where  $k=0,1,2,\cdots$, and $\tau\neq 0$ is hoped to converge to a root of $h_f$. Suppose the iteration of (37) is convergent to  $\omega^{*} $, a root of the IF equation and  a  function of  $(t, \omega_{0})$,  then we have 
\begin{align}
	h_f(t, \omega^{*})=0. \nonumber
\end{align}

To show the TF content  of the signal more accurately and with better readability,   we reassign  the TF value of ($t, \omega_0$) to the location  ($t, \omega^{*}$) as follows: 
\begin{equation}
	{T_r}(t,\omega^*):=S_f^g(t,\omega_{0})+ S_f^g(t,\omega^*). 
\end{equation}
Similar to the expression (9) in the SST, this reassignment operation above  can be formulated as 
\begin{equation}
	{T_r}(t,\omega)=\int_{|S_f^g(t,\eta)|>\lambda}S_f^g(t,\eta)\delta (\omega-{\omega^*}(t,\eta))\mathrm{d}\eta.
\end{equation} 
Due to the relocation of  the TF points only along the frequency direction, the original signal can be approximated  by 
\begin{equation}
	f(t)\approx\frac{1}{g(0)} \int {T_r}(t,\omega)\mathrm{d}\omega.
\end{equation} 

In the following, we utilize two  concrete  IF equations, i.e., the $IF_1$ and $IF_2$ of the chirps,  to discuss the  convergence of the fixed point algorithm (37). For the  $IF_1$,  the iteration of  $\omega$ at time $t$   can be   expressed as 
\begin{align}
	\omega_{k+1}=	\omega_{k}+\tau \left|  \frac{\frac{\partial}{\partial \omega}S_{f}^g(t, \omega)|_{\omega=\omega_k}}{S_{f}^g(t, \omega_k)} \right|^2  
	= 	\omega_{k}+ \tau\frac{\sigma^4(\omega_k-\phi'(t))^2}{{1+\sigma^4c^2}}. 
\end{align}
Expression  (41) yields that 
\begin{align}
	&	|\omega_{k+1}-\omega_{k}|\nonumber\\&=	\left| \omega_{k}-\omega_{k-1}+\frac{\tau\sigma^4}{{1+\sigma^4c^2}}((\omega_k-\phi'(t))^2-(\omega_{k-1}-\phi'(t))^2)\right| \nonumber\\
	&\leq  \left| 1+\frac{\tau\sigma^4}{{1+\sigma^4c^2}}(\omega_{k}+\omega_{k-1}-2\phi'(t))\right|\left|\omega_{k}-\omega_{k-1}\right|.  
\end{align}
The parameter $\tau$ can be set as $0<\tau<\frac{1+\sigma^4c^2}{\sigma^4\tilde{\Delta}}$  if  the  initial point satisfies $\omega_{0}<\phi'(t)$, and set as $-\frac{1+\sigma^4c^2}{\sigma^4\tilde{\Delta}}<\tau<0$ if the initial point satisfies $\omega_{0}>\phi'(t)$, to ensure the convergence. Here $\tilde{\Delta}$ denotes the  support of $\mathrm{e}^{-\frac{\sigma^2}{2(1+\sigma^4c^2)}\omega^2}.$

When considering the  $IF_2$, the iteration of (37)  can be  further written  as 
\begin{align}
	\omega_{k+1}=	\omega_{k}+\tau \Re\left\lbrace  \frac{\frac{\partial}{\partial \omega}S_{f}^g(t, \omega)|_{\omega=\omega_k}}{S_{f}^g(t, \omega_k)} \right\rbrace 
	= 	\omega_{k}- \tau\frac{\sigma^2(\omega_k-\phi'(t))}{{1+\sigma^4c^2}}, 
\end{align}
From (43), we have 
\begin{align}
	&	|\omega_{k+1}-\omega_{k}| \nonumber \\&=	\left| \omega_{k}-\omega_{k-1}-\frac{\tau\sigma^2}{{1+\sigma^4c^2}}(\omega_k-\phi'(t)-(\omega_{k-1}-\phi'(t)))\right| \nonumber\\
	&\leq  \left| 1-\frac{\tau\sigma^2}{{1+\sigma^4c^2}}\right|\left|\omega_{k}-\omega_{k-1}\right|.  
\end{align}
Therefore,  iteration (43) is convergent if $\tau$ is taken as a value such that  $|1-\frac{\tau\sigma^2}{\sqrt{1+\sigma^4c^2}}|<1$.

It should be pointed out that the multi-synchrosqueezing transform [44] is a  specific case of the  IF equation-based reassignment method. Indeed, when we consider the  IF equation (32) and take the value of $\tau$ in (37) as -1, then  iteration (37)  can be written  as 
\begin{align}
	\omega_{k+1}=	\hat{\omega}_1(t,\omega_k), ~ k=0,1,2,\cdots,
\end{align}
which corresponds to the multiple SST operation in [44]. Beyond this  specific  case,  we can consider selecting another value of $\tau$  in (37) to further improve the convergence.  

Moreover, we can also use the Newton algorithm to get a root of $h_f(t,\omega)$.  For  the initial point ($t, \omega_0$)  in $ \Omega$,    the Newton iteration formula can be expressed as  
\begin{align}
	\omega_{k+1}&=\omega_k-\frac{h_f(t,\omega_k)}{\frac{\partial}{\partial \omega}h_{f}(t, \omega)|_{\omega=\omega_k}}, k=0,1,2,\cdots.
\end{align}
Specifically,   with regard to the case of the $IF_1$ of  chirp signals, Newton iteration (46) can be further  written as
\begin{align}
	\omega_{k+1}&=\frac{1}{2}(\omega_{k}+\phi'(t)),
\end{align}
which leads to 
\begin{align} 
	|\omega_{k+1}-\omega_{k}|=\frac{1}{2}\left|\omega_{k}-\omega_{k-1}\right|.  \nonumber
\end{align} 
It is easy to prove  that the  Newton iteration used for solving  the IF equation $IF_1$ is convergent. 

Due to that 
\begin{equation}
	h_f(t,\omega)=	\left| \frac{\frac{\partial}{\partial \omega}S_{f}^g(t, \omega)}{S_{f}^g(t, \omega)} \right|^2, \text{~~for} ~(t, \omega)\in \Omega,  \nonumber
\end{equation}
we have 
\begin{equation}
	\begin{split}
		&\frac{h_f(t,\omega)}{\frac{\partial}{\partial \omega}h_{f}(t, \omega)}=\\& \frac{ |{\frac{\partial}{\partial \omega}S_{f}^g(t, \omega)}||{S_{f}^g(t, \omega)} |}{2(\frac{\partial}{\partial \omega}|{\frac{\partial}{\partial \omega}S_{f}^g(t, \omega)}||S_{f}^g(t, \omega)|-\frac{\partial}{\partial \omega}|S_{f}^g(t, \omega)||{\frac{\partial}{\partial \omega}S_{f}^g(t, \omega)}|)}.
	\end{split}
\end{equation}
Since 	$|{\frac{\partial}{\partial \omega}S_{f}^g(t, \omega)}|=\left| {S_{f}^{tg}(t, \omega)}\right|$,  $\frac{\partial}{\partial \omega}|S_{f}^g(t, \omega)|=\Re \left\lbrace  \frac{-jS_{f}^{tg}(t, \omega)}{S_{f}^g(t, \omega)}\right\rbrace |S_{f}^g(t, \omega)|$, and $\frac{\partial}{\partial \omega}|{\frac{\partial}{\partial \omega}S_{f}^g(t, \omega)}|=\frac{\partial}{\partial \omega}|S_{f}^{tg}(t, \omega)|$$=\Re \left\lbrace \frac{-jS_{f}^{t^2g}(t, \omega)}{S_{f}^{tg}(t, \omega)}\right\rbrace |S_{f}^{tg}(t, \omega)|$, then the Newton iteration  can be more efficiently computed by
\begin{align}
	\omega_{k+1}&=\omega_k-\frac{1}{2\left( \Re\left\lbrace \frac{-jS_{f}^{t^2g}(t, \omega_k)}{S_{f}^{tg}(t, \omega_k)}\right\rbrace-\Re \left\lbrace  \frac{-jS_{f}^{tg}(t, \omega_k)}{S_{f}^g(t, \omega_k)}\right\rbrace \right)  }.
\end{align}

In order to  improve the convergence of the Newton algorithm when used for $IF_1$, we can reformulate (46) as 
\begin{align}
	\omega_{k+1}&=\omega_k-2\frac{h_f(t,\omega_k)}{\frac{\partial}{\partial \omega}h_{f}(t, \omega)|_{\omega=\omega_k}}. \label{eq45}
\end{align}
One iteration of (50)  can converge to the solution of the  IF equation of chirps. Therefore,  developing an effective iterative algorithm is another powerful way to obtain a concentrated  TF distribution, compared with the IF estimator. 

 It is noted that the reassignment operation along the frequency direction is not sufficient for GD characterization when addressing the mixed signals. Future works will consider new reassignment operations with an adaptive pattern to yield a high-accuracy TF description.  

\subsection{Comparison}

This section uses a simulated example to test the performance of the  IF equation-based TF post-processing methods and compare them with the IF estimator-based TF post-processing methods.

\begin{figure}[!b]
	\centering
	\begin{minipage}{0.84\linewidth}
		\centerline{\includegraphics[width=1\textwidth]{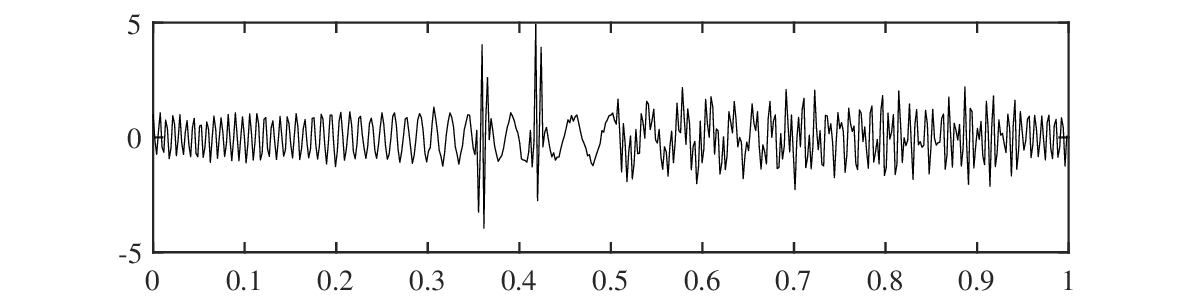}}
		\centerline{(a)}
	\end{minipage}
	\begin{minipage}{0.485\linewidth}
		\centerline{\includegraphics[width=1\textwidth]{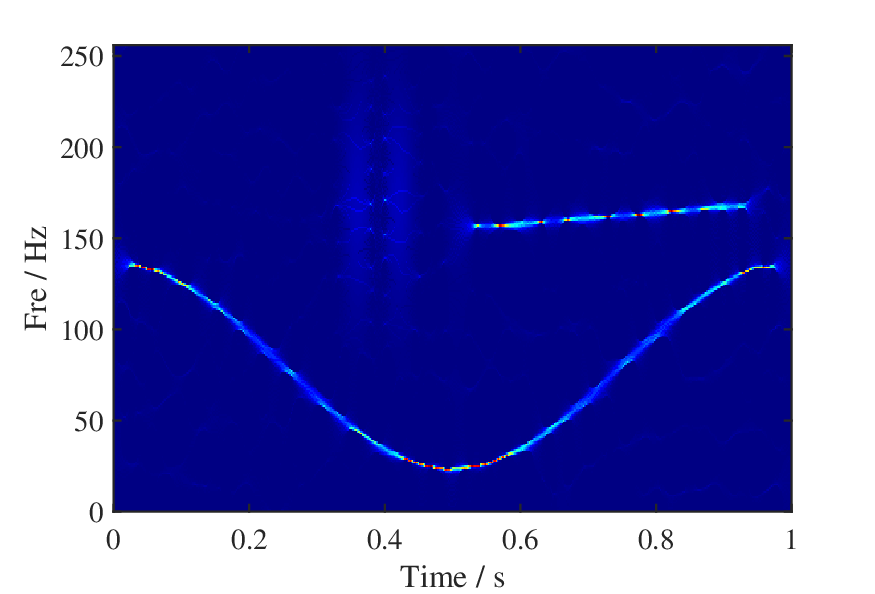}}
		\centerline{(b)}
	\end{minipage}
	\begin{minipage}{0.485\linewidth}
		\centerline{\includegraphics[width=1\textwidth]{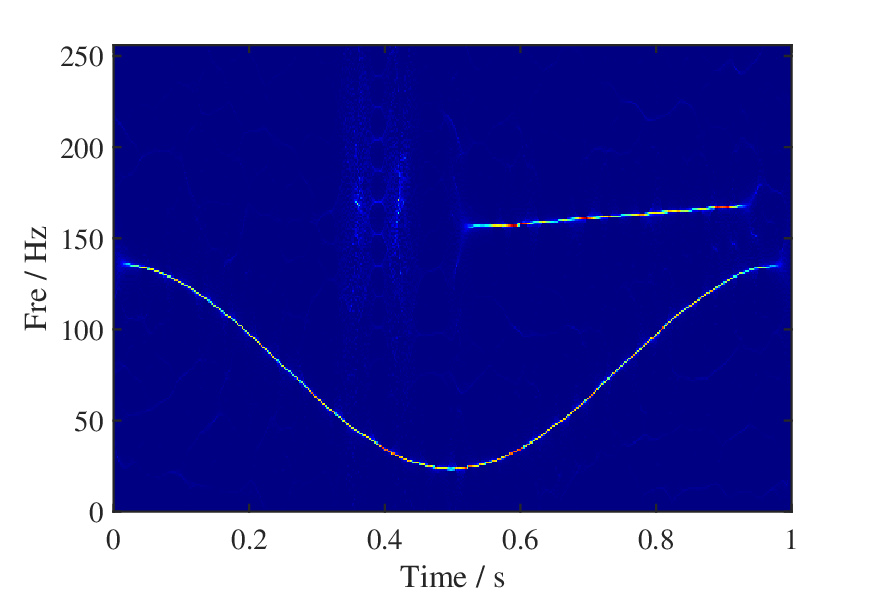}}
		\centerline{(c)}
	\end{minipage}
	\begin{minipage}{0.485\linewidth}
		\centerline{\includegraphics[width=1\textwidth]{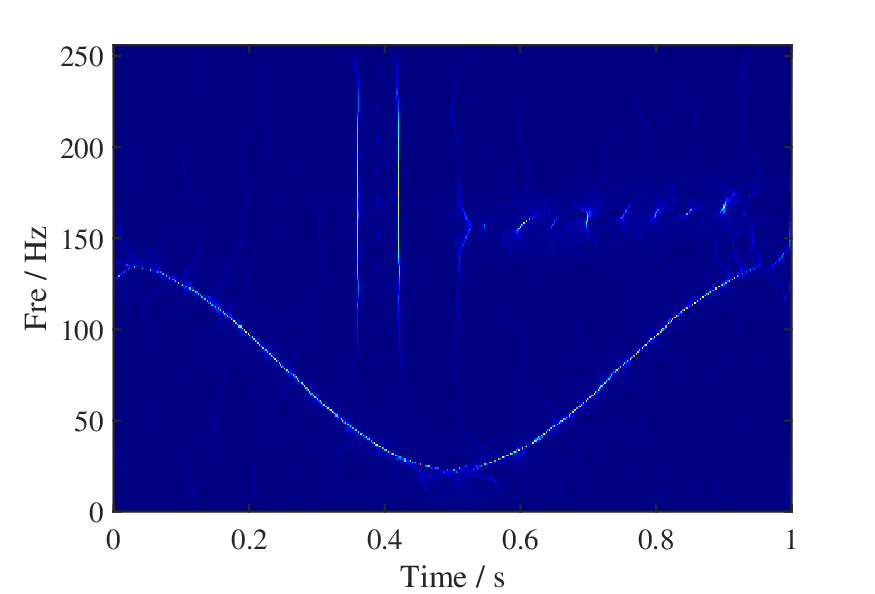}}
		\centerline{(d)}
	\end{minipage}
	\begin{minipage}{0.485\linewidth}
		\centerline{\includegraphics[width=1\textwidth]{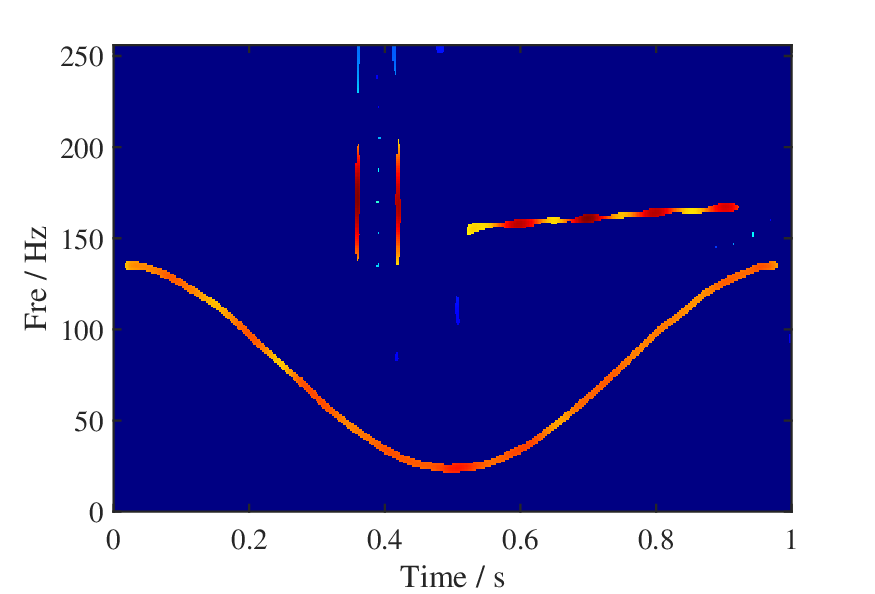}}
		\centerline{(e)}
	\end{minipage}
	\begin{minipage}{0.485\linewidth}
		\centerline{\includegraphics[width=1\textwidth]{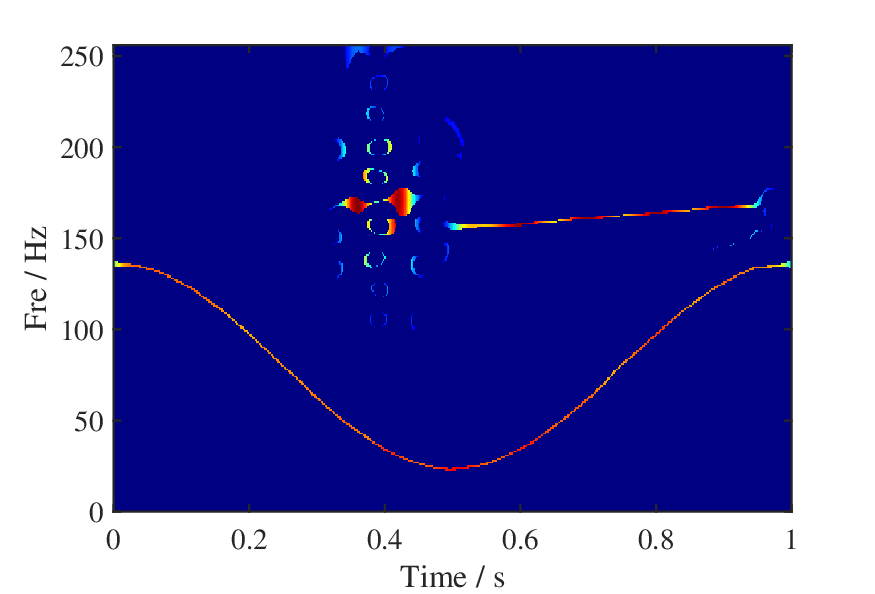}}
		\centerline{(f)}
	\end{minipage}
	\begin{minipage}{0.485\linewidth}
		\centerline{\includegraphics[width=1\textwidth]{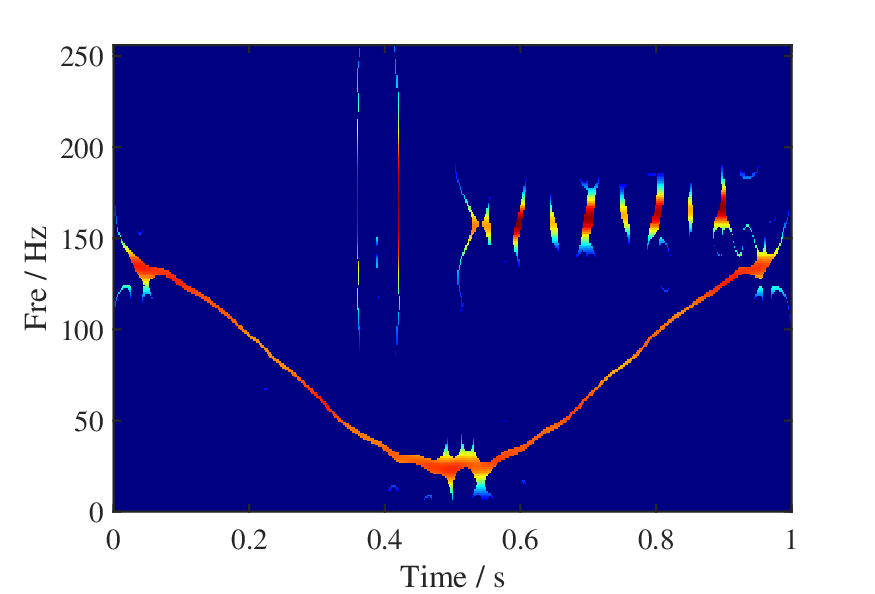}}
		\centerline{(g)}
	\end{minipage}
	\begin{minipage}{0.485\linewidth}
		\centerline{\includegraphics[width=1\textwidth]{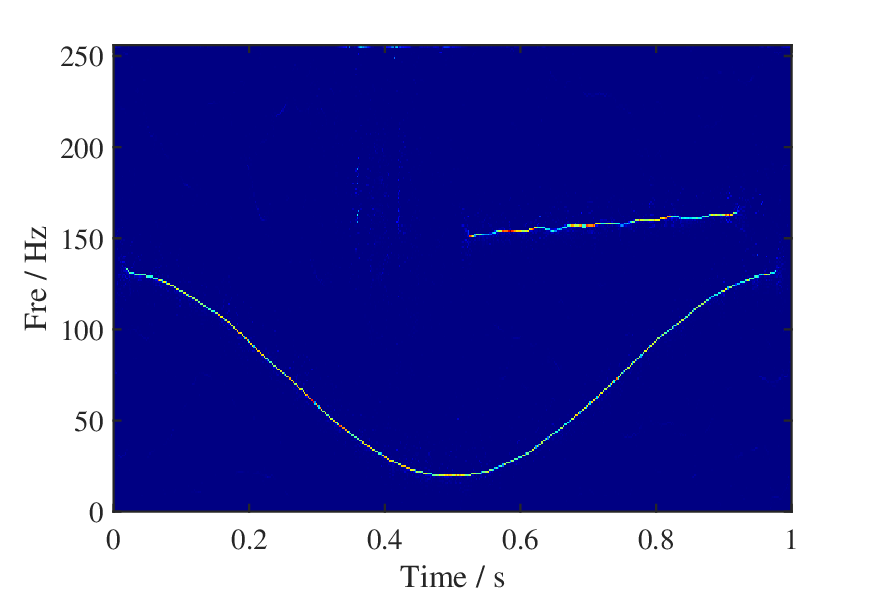}}
		\centerline{(h)}
	\end{minipage}
	\begin{minipage}{0.485\linewidth}
		\centerline{\includegraphics[width=1\textwidth]{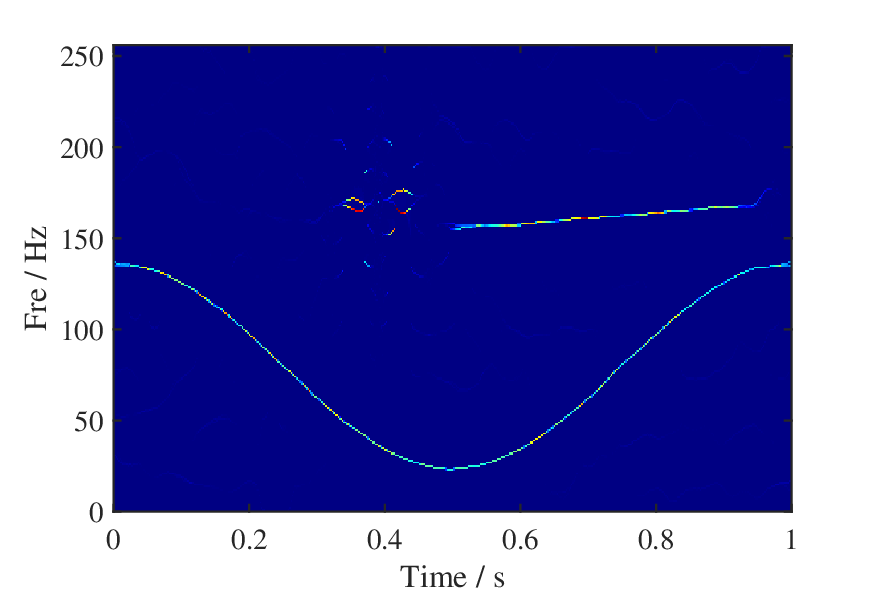}}
		\centerline{(i)}
	\end{minipage}
	\caption{ \small Evaluation of TF results on a known signal structure. (a) Waveform of the test signal, (b) TF  plot by FSST,  (c)  TF plot by SSST, (d)  TF plot by TSST,   (e)  TF plot by ETIF$_1$, (f)  TF plot by ETIF$_2$, (g)  TF plot by ETIF$_3$, (h) TF plot by RMNIF$_1$, (i) TF plot by RMFPIF$_2$.}
\end{figure}
The waveform of the test signal is illustrated in  Fig. 1(a), which contains two impulsive signals, a chirp, and a cosine modulation component. We compute the TF representations of the signal using the FSST [24], second-order SST (SSST for short) [42], time-reassigned SST (TSST for short) [46],  extracting transform based on $IF_i$ (ETIF$_i$ for short, $i=1,2,3$),   reassignment method based on the Newton iteration and  $IF_1$ (RMNIF$_1$ for short), and reassignment method based on the fixed point algorithm and $IF_2$ (RMFPIF$_2$ for short).  From the results, we can observe that the FSST and SSST yield good energy concentration for the chirp and cosine-modulated signal but fail to characterize the impulsive signals effectively. The TSST provides a better TF localization for the transient signals while degrading the TF sharpness of the stationary mode.  For our introduced extracting transforms, the ETIF$_1$  provides excellent performance in dealing with the mixed signal; ETIF$_2$  provides a high-concentration TF representation for the chirp and the cosine modulation signal but results in a blurred TF representation for impulsive signals; ETIF$_3$ leads to the opposite result compared with ETIF$_2$. The results imply the fact that the IF equation $IF_2$ is more suitable for characterizing the time-varying signals, like chirps and cosine modulation signals, whereas the  $IF_3$ is more useful in detecting the impulsive signals. The RMNIF$_1$  and RMFPIF$_2$ have better TF plots than FSST and achieve competitive results with the SSST. Unfortunately, all of the reassignment methods considered here lead to the blurred characterization of the impulsive signals because of the limitation of the reassignment operation along the frequency direction, as shown in Fig. 1(b, c, h, i).  In this regard,  the extracting transform appears to be more effective.

Although the ETIF$_1$ shows excellent potential in handling the mixed signal,  there still exist two issues to be addressed. First,  ETIF$_1$ does not achieve an ideal energy concentration, still resulting in TF representation diffusion because Rényi entropy of the TF representation by ETIF$_1$ is 11.52, larger than the values by  SSST (11.17) and ETIF$_2$ (10.88).  Second, it is subject to the extracting operation and the function of the  IF equation,  sometimes leading to discontinuous  TF information curves when pursuing a high-concentration TF representation. Fig. 2(a) displays the TF result of  ETIF$_1$ when using a relatively small threshold value in (35). Obviously, the disconnected characterization of the chirp occurs in the TF  plot, as marked with the pink box. 

\begin{figure}[!tbh]
	\centering
	\setlength{\belowcaptionskip}{-0.2cm}
	\begin{minipage}{0.485\linewidth}
		\centerline{\includegraphics[width=1\textwidth]{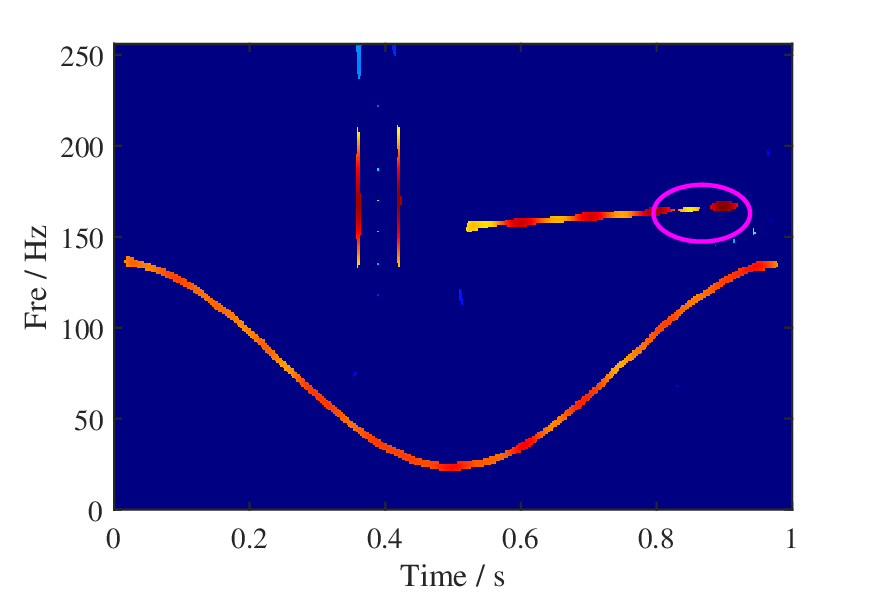}}
		\centerline{(a)}
	\end{minipage}
	\begin{minipage}{0.485\linewidth}
		\centerline{\includegraphics[width=1\textwidth]{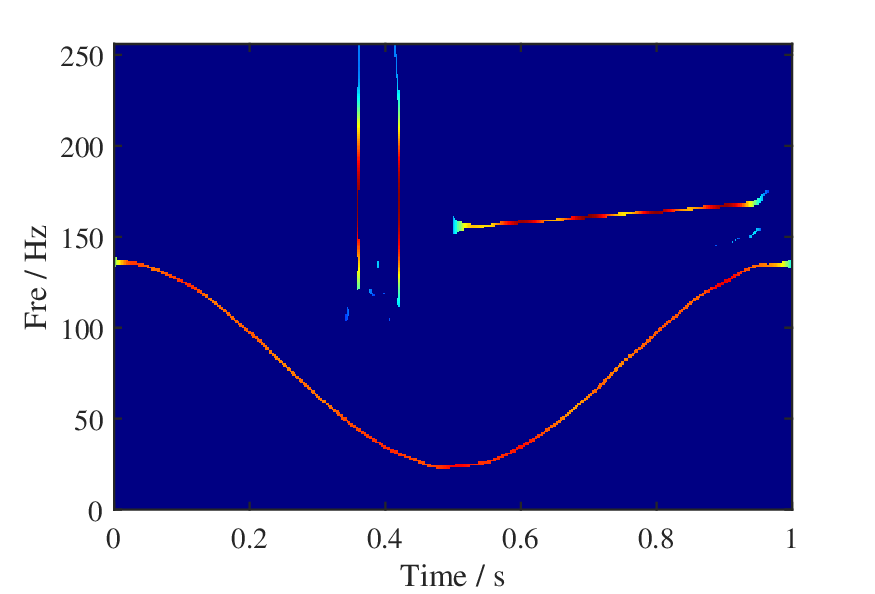}}
		\centerline{(b)}
	\end{minipage}
	\caption{ \small Evaluation of TF results on a known signal structure. (a) TF plot by ETIF$_1$, (b)  TF plot by ETCIFLS.}
\end{figure}

\section{Combination of IF equations based on local sparsity}

To address the two issues above, this section introduces a novel approach to combine different IF equations to minimize energy spreading based on local sparsity. 

From the analysis in the previous section, we know that the $IF_2$ provides a high-concentration TF representation for time-varying signals. Namely, it guarantees the sparsity of the TF distribution when characterizing time-varying signals.  Alternatively,  the $IF_3$ leads to a sparse TF representation for frequency-varying signals.  The attempt is to select the sparser representation from the TF distributions using the two IF equations, i.e.,  $IF_2$ and $IF_3$.  

We here consider a discrete signal with $N$ samples and suppose $IF_2$ and $IF_3$  be known, which leads to two  TF representations of ETIF$_2\in \mathbb{C}^{M\times N}$ and ETIF$_3 \in \mathbb{C}^{M\times N}$. For each TF bin $(m,n)$, a local TF representation  performed by a two-dimensional window $W_s$  with $(2n_w+1)\times (2n_w+1)$ elements can be obtained that 
\begin{equation}
	\widetilde{\text{ETIF}}^{m,n}_i:=\widetilde{\text{ETIF}_i}[m-n_w:m+n_w,n-n_w:n+n_w]\circledast W_s,	
\end{equation}
where $\widetilde{\text{ETIF}_i}\in \mathbb{C}^{(M+2n_w)\times (N+2n_w)}$ is the extension of  ETIF$_i$ by zero padding, $i=2, 3$,   $m=n_w+1, n_w+2,\cdots, n_w+M$, $n=n_w+1, n_w+2,\cdots, n_w+N$,  and $\circledast$ denotes the element-wise product.  The sparsity of each  TF region  $\widetilde{\text{ETIF}}^{m,n}_i$ is then measured via calculating  the  Gini index [58] as  
\begin{equation}
	G(X)=1-2\sum_{i=1}^{N_s}\frac{x_i}{\|\vec{x}\|_1}\left( \frac{N_s-i+\frac{1}{2}}{N_s}\right),	
\end{equation}
where $\vec{x}=[x_1~ x_2 \cdots x_{N_s}]$ containing the elements of the magnitude of the input matrix
$X$ in ascending order, and $\|\vec{x}\|_1$ denotes the $l_1$-norm of vector $\vec{x}$. 
The value of the Gini index is limited to the interval [0, 1], where the value near  0 denotes a group of samples with equally distributed energy and near 1 means maximum energy concentration. 

After computing the local sparsity of all local representations  using  $IF_2$ and $IF_3$, a TF representation is constructed to indicate which exhibits the highest local sparsity for each TF bin: 
\begin{equation}
	{\text{ETCIFLS}}[{m, n}]:=\text{ETIF}_p[m, n],	
\end{equation}
where $p=\max_i G(\widetilde{\text{ETIF}}^{m,n}_i)$. This representation is a combination of extracting transform with different IF equations and is an optimum representation in terms of local sparsity. We denote it as ETCIFLS.

Fig. 2(b) depicts the TF representation of the test signal in  Fig. 1(a)  using ETCIFLS.  Rényi entropies of the two  TF representations in Fig. 2 are computed as  11.69 (ETIF$_1$) and  10.79 (ETCIFLS). The experimental results show that the energy concentration of ETCIFLS  is greatly improved compared to that of the ETIF$_1$, and  ETCIFLS  provides more detailed and accurate TF information than ETIF$_1$.

It is worth noting that a disadvantage of this method, in comparison to the approach presented in the previous section, is that the  ETCIFLS  requires much higher computational resources because it involves not only the STFT calculations with different windows but also the local sparsity detections. When considering the signal of  $N$ samples, the STFT can be implemented by the FFT, which requires $O(N^2\log_2N)$  operations, and the sparsity measure requires $O(N^2(2n_w+1)^2)$ computations. Therefore, the total computing complexity of  ETCIFLS  is $O(N^2(\log_2N+(2n_w+1)^2))$, in which, a large one of $n_w$  inevitably leads to an increase in time cost.  

The window $W_s$, responsible for the local sparsity computation, is designed as a two-dimensional Hamming window. The dimensions of $W_s$  are related to the resolutions of ETIF$_1$ and ETIF$_2$. A large size generally increases the computation, whereas a small size easily results in a  lack of TF information. Through experiments,  we suggest the value of $n_w$ of   $W_s$  can be considered as  [10, 35], which leaves this window enough room for including the frequency components present in all TF representations and usually produces a   satisfactory result.  

\section{Numerical validation}

In this section, we conduct numerical simulations to illustrate the effectiveness of the proposed methods.  For comparison, the state-of-the-art highly concentrated schemes, namely the SET [39], the SSST [42],   the second-order synchroextracting wavelet transform (SSEWT for short)  [7],  and the MDD [59] have been used.  

\subsection{Simulated signal}
We first use the proposed  methods  to  detect a mixed signal composed of three harmonic-like  components and two impulsive-like  components, which is given by 
\begin{equation}
	\begin{split}
		&f(t)=f_1(t)+f_2(t)+f_3(t)+f_4(t)+f_5(t)+n(t), \\
		&f_1(t)=\sin(40\pi  t),     ~ t\in [0.33, 1.4],\\
		&f_2(t)=\sin(80\pi t),     ~ t\in [0.33, 1.4],\\
		&f_3(t)=\sin(120\pi t+\pi t^2)),     ~ t\in [0.33, 1.4],\\
		&f_4(t)=\Re\{FT(0.02\exp(40\pi j t))\}, ~  t\in [0, 2],\\
		&f_5(t)=\Re\{FT(0.02\exp(70\pi j t ))\}, ~  t\in [0, 2], 
	\end{split}
\end{equation}
where $n(t)$ is the Gaussian  noise  with  the $\text{SNR}=6$ dB, and $FT$ denotes the Fourier transform.   The sampling frequency is  256  Hz.

Fig. 3 shows the TF representations obtained by the SET, SSST, SSEWT, MDD, ETIF$_1$, and ETCIFLS. It can be seen from the results that, the SET and SSST  give clear TF representations for  $f_1(t)$, $f_2(t)$, and $f_3(t)$, but fail to localize the other two impulsive components effectively. In contrast,  the SSEWT can produce a concentrated TF representation for the transient signals but provides erroneous TF content for the harmonic models. The TF representation by MDD is blurred and there are heavy cross-terms between the neighboring modes in the TF plane.  The ETIF$_1$ and  ETCIFLS  are superior to others,  yielding good characterization of both time-varying harmonic and frequency-varying harmonic components.  It should be noted that  ETIF$_1$ gives discontinuous TF ridges, and misses a small amount of boundary information when pursuing a high-concentration TF representation.  To show the performance of the proposed approaches quantitatively,  we compute the time cost and  Rényi entropy (RE) of the six TF analysis methods.  The tested computer configuration is as follows: Intel Core  i9-11900KF 3.50  GHz, 32.0 GB of RAM, and MATLAB version R2021a. 
As presented in  Table \uppercase\expandafter{\romannumeral1},  the SSEWT and ETCIFLS  have smaller values of RE  among all the selected TF representations, so they have better energy concentration.  The ETIF$_1$ takes less time than SSST and MDD to represent the signal. The ETCIFLS  is relatively more time-consuming than the others to produce the result. 
\begin{figure}[pbth!]
	\vspace{-0.1cm}
	\setlength{\belowcaptionskip}{-0.2cm}
	\centering
	\begin{minipage}{0.485\linewidth}
		\centerline{\includegraphics[width=1\textwidth]{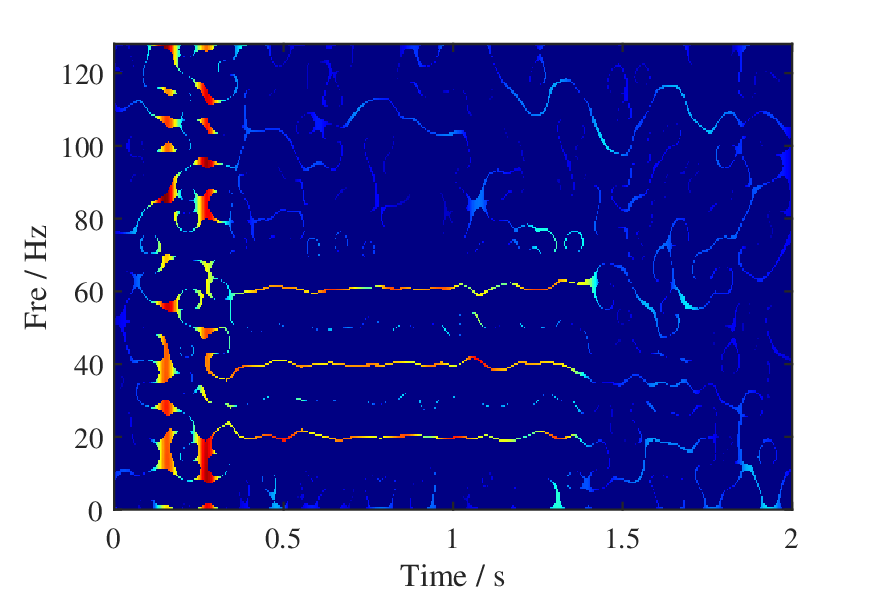}}
		\centerline{(a)}
	\end{minipage}
	\begin{minipage}{0.485\linewidth}
		\centerline{\includegraphics[width=1\textwidth]{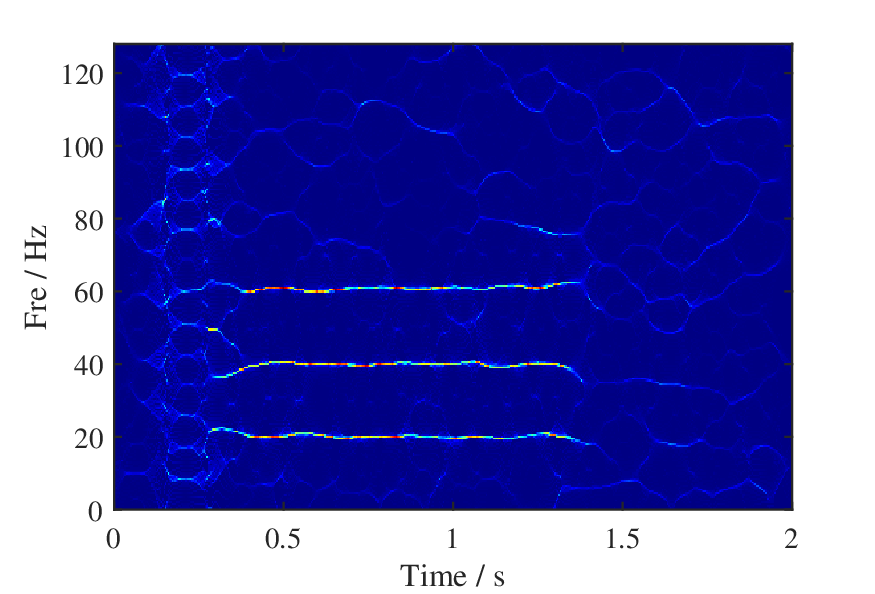}}
		\centerline{(b)}
	\end{minipage}\\ \begin{minipage}{0.485\linewidth}
		\centerline{\includegraphics[width=1\textwidth]{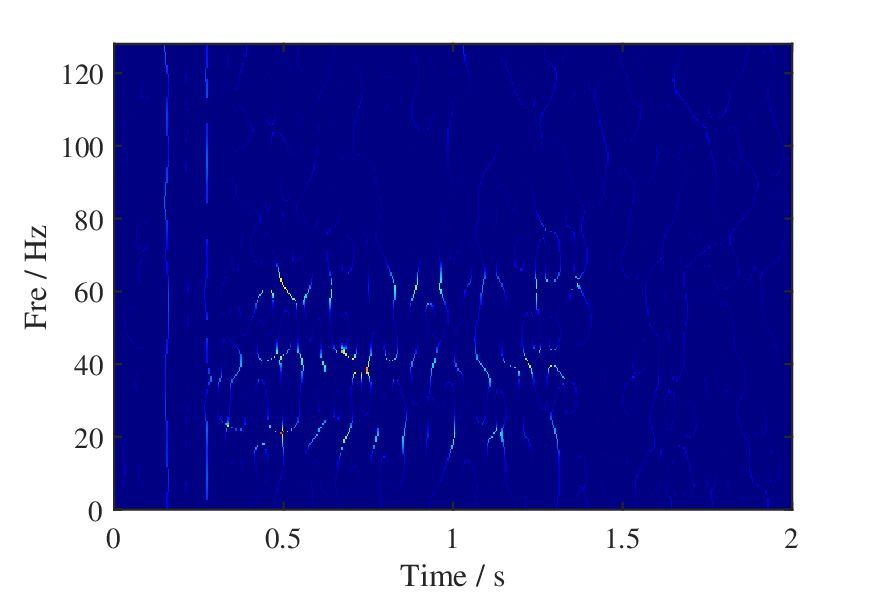}}
		\centerline{(c)}
	\end{minipage}
	\begin{minipage}{0.485\linewidth}
		\centerline{\includegraphics[width=1\textwidth]{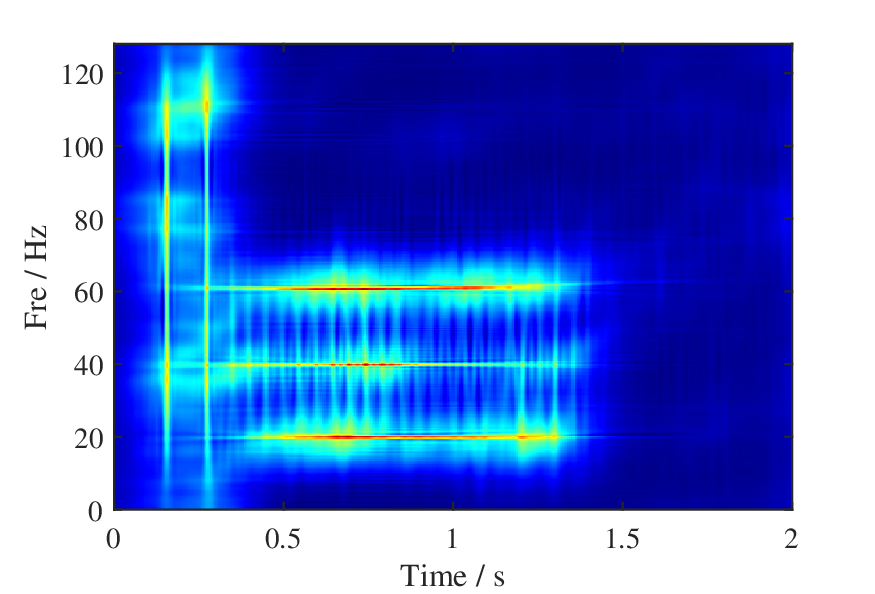}}
		\centerline{(d)}
	\end{minipage}
	\begin{minipage}{0.485\linewidth}
	\centerline{\includegraphics[width=1\textwidth]{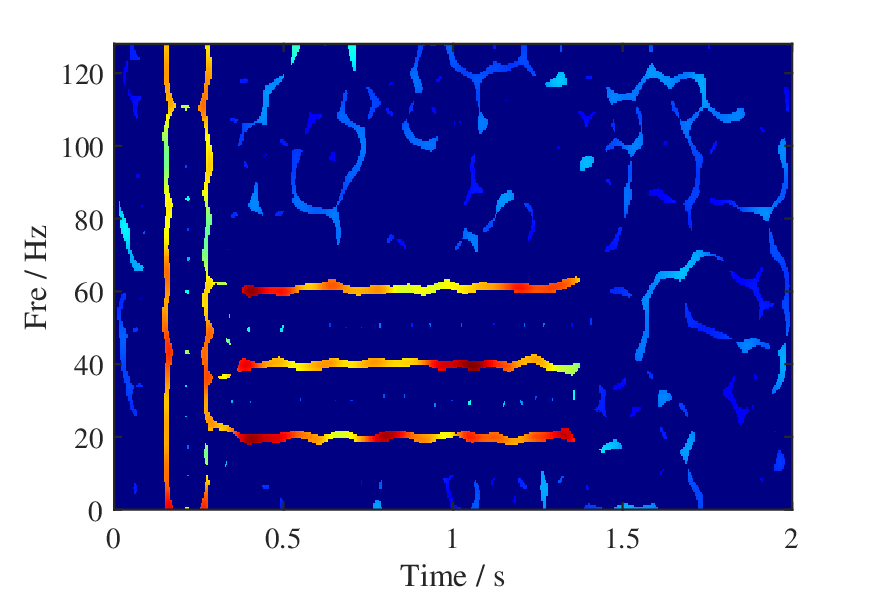}}
	\centerline{(e)}
\end{minipage}
	\begin{minipage}{0.485\linewidth}
	\centerline{\includegraphics[width=1\textwidth]{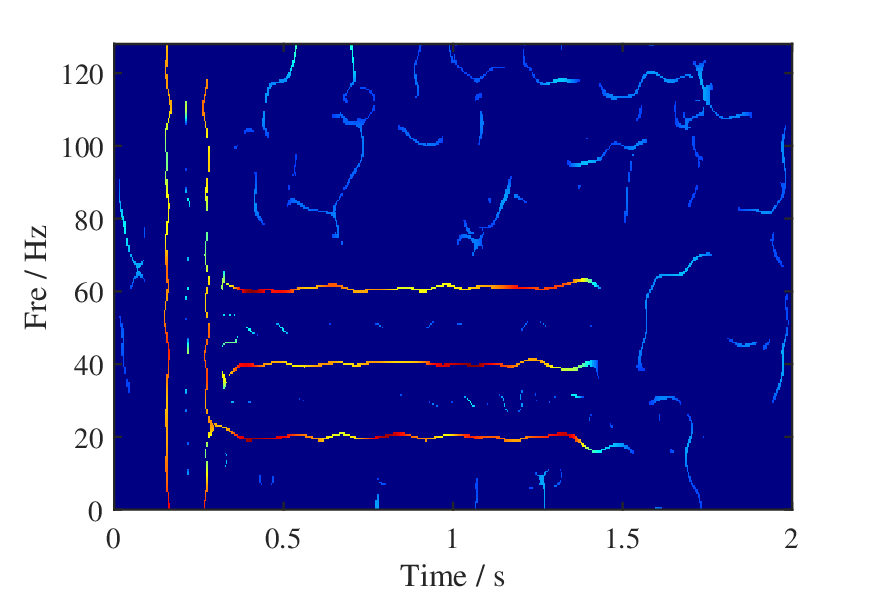}}
	\centerline{(f)}
\end{minipage}
	\caption{ \small  TF results of signal (54) using various analysis  methods.  (a)  SET,  (b)  SSST, (c) SSEWT, (d) MDD (1, 0.003), (e) ETIF$_1$, (f) ETCIFLS ($n_w=41$).}
\end{figure}

\begin{table}[!tbph]
	\vspace{0cm}
	\caption{ \label{tab4} Performance comparison of the selected TF representations  for  signal (54). } \centering
	\begin{tabular}{ccccccccccc}
		\Xhline{.6pt}  Method&{SET}&{SSST}&{SSEWT}&{MDD}&{ETIF$_1$}&{ETCIFLS}\\
		\hline
		Time (s)&0.041 &0.219&0.062&0.268 &0.035&5.951\\
		RE  &  12.417 &  12.707  & 11.115  & 16.596 &  13.037  & 11.520\\	
		\Xhline{.6pt}  
	\end{tabular}
	\setlength{\belowcaptionskip}{-0.15cm}
\end{table}

We next consider a four-component AM-FM signal including  a signal with hyperbolic
	frequency modulation, a cosine  modulation  signal,  and   two impulse components, that is 
\begin{equation}
	\begin{split}
		f(t)&=f_1(t)+f_2(t)+f_3(t)+f_4(t)+n(t), \\
		f_1(t)&=\cos(20\pi (15t+5\log(|t|)),  ~  t\in (0, 1], \\
		f_2(t)&=(1.2+0.1\sin(10\pi t))\sin(2\pi(145t-4\times\\&\exp(-0.3t+0.4)\sin(6\pi(t2-0.2)))),~  t\in [0.2, 0.5],\\
		f_3(t)&=5\exp(-10000\pi(t-0.74)^2)\cos(280\pi t),  ~  t\in [0, 1],\\
		f_4(t)&=5\exp(-10000\pi(t-0.8)^2)\cos(280\pi t),   ~  t\in [0, 1],
	\end{split}
\end{equation}
where  $n(t)$ is the Gaussian noise with the SNR$=5$ dB.   The sampling frequency is 512 Hz. 

\begin{figure}[pbt!]
	\vspace{-0.1cm}
	\setlength{\belowcaptionskip}{-0.2cm}
	\centering
	\begin{minipage}{0.485\linewidth}
		\centerline{\includegraphics[width=1\textwidth]{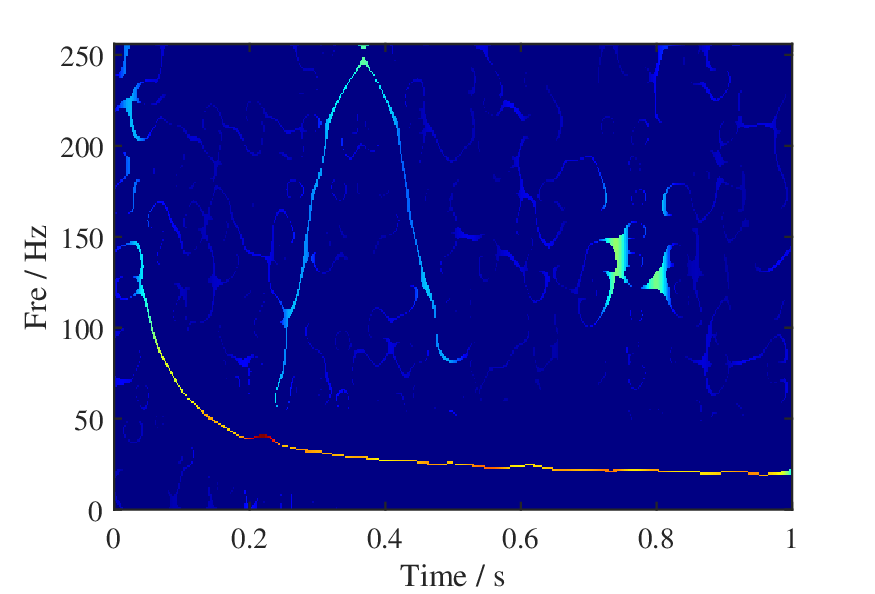}}
		\centerline{(a)}
	\end{minipage}
	\begin{minipage}{0.485\linewidth}
		\centerline{\includegraphics[width=1\textwidth]{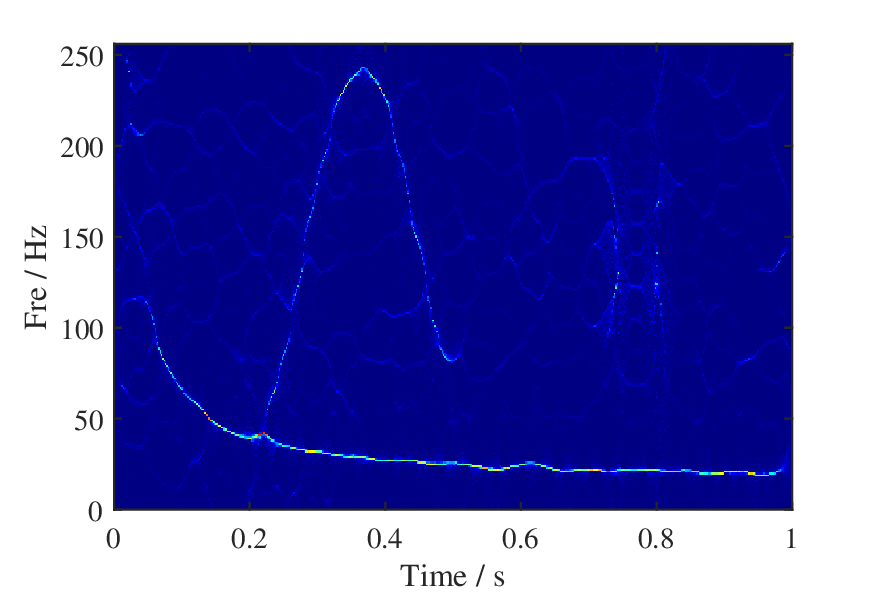}}
		\centerline{(b)}
	\end{minipage}\\ \begin{minipage}{0.485\linewidth}
		\centerline{\includegraphics[width=1\textwidth]{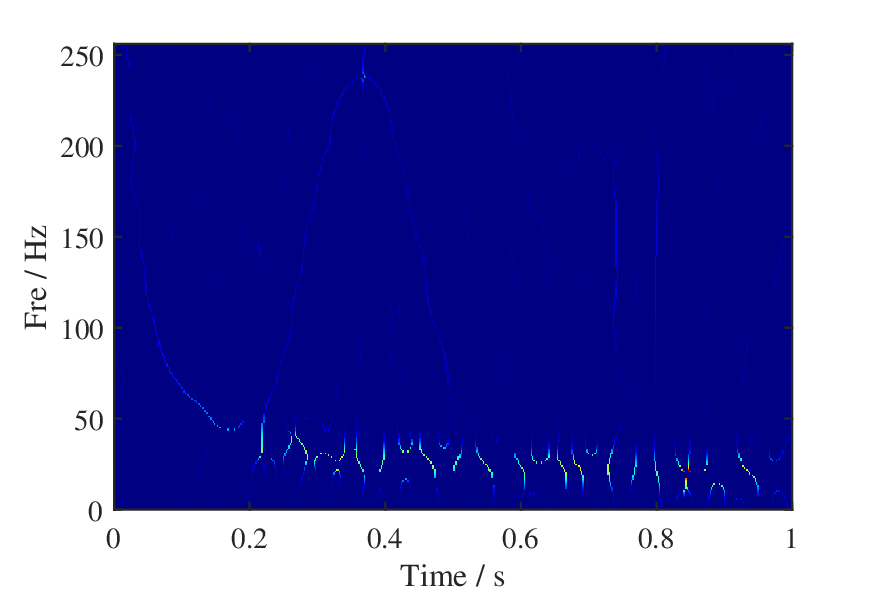}}
		\centerline{(c)}
	\end{minipage}
	\begin{minipage}{0.485\linewidth}
		\centerline{\includegraphics[width=1\textwidth]{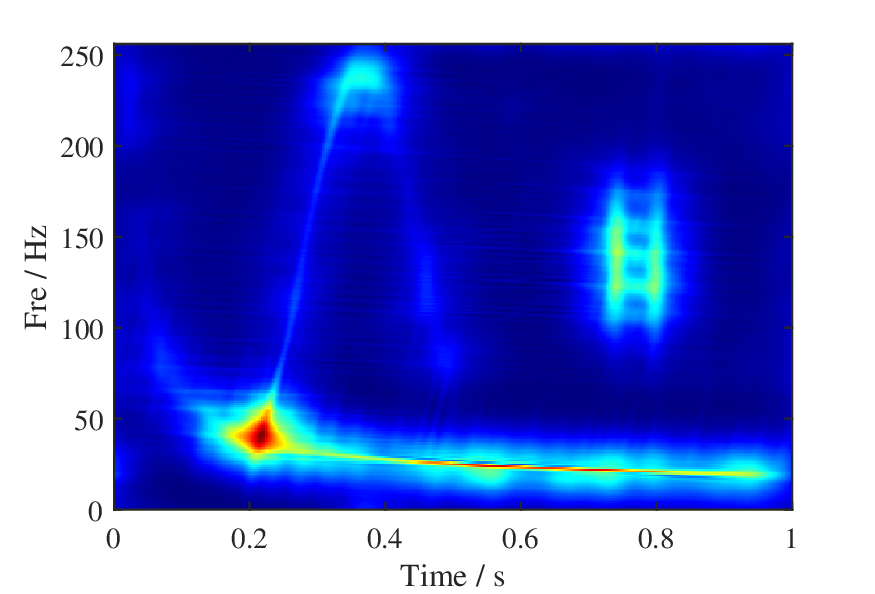}}
		\centerline{(d)}
	\end{minipage}
	\begin{minipage}{0.485\linewidth}
		\centerline{\includegraphics[width=1\textwidth]{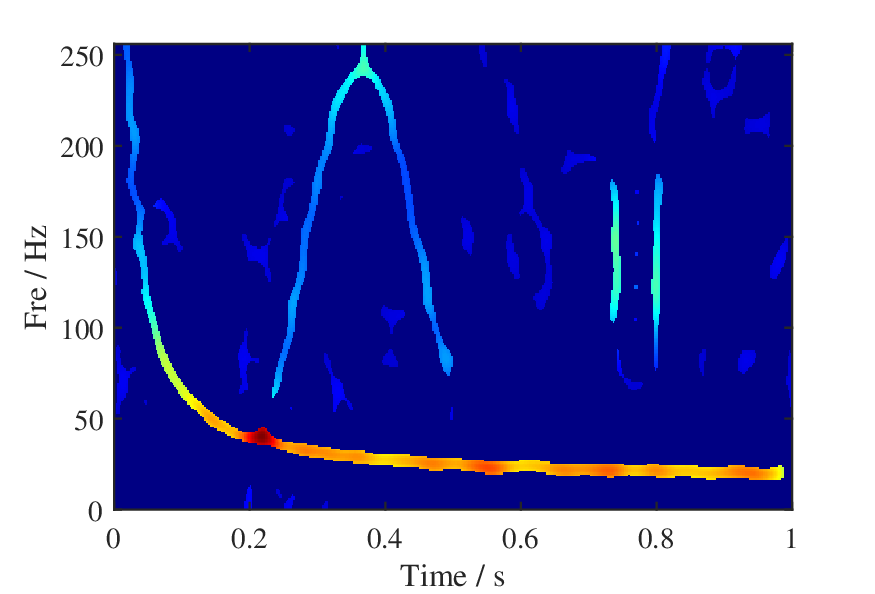}}
		\centerline{(e)}
	\end{minipage}
	\begin{minipage}{0.485\linewidth}
		\centerline{\includegraphics[width=1\textwidth]{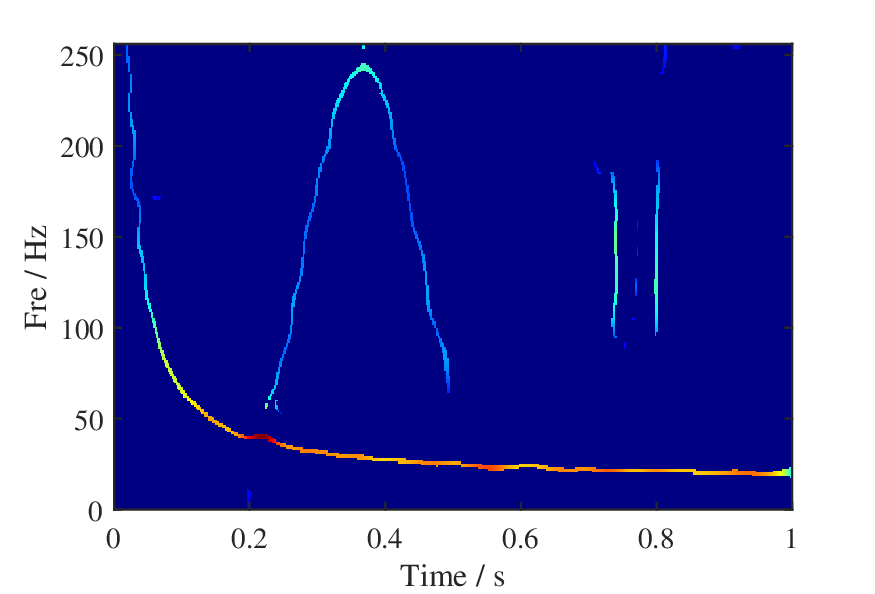}}
		\centerline{(f)}
	\end{minipage}
	\caption{ \small  TF results of signal (55) using various analysis  methods.  (a)  SET,  (b)  SSST, (c) SSEWT, (d) MDD (1, 0.003), (e) ETIF$_1$, (f) ETCIFLS ($n_w=41$).}
\end{figure}

The TF representations of the test signal (55) using the six TF methods are provided in Fig. 4.  The results show  SET and SSST are ineffective in characterizing the impulsive-like models, from which, it is easy to recognize some signal information as noise. The SSEWT  provides a better TF localization for the transient signals while degrading the TF sharpness of the stationary part of the hyperbolic signals (see TF plot in the time interval [0.2, 1]). As given in Fig. 4(d), the MDD has a blurred TF representation, from which it is difficult to identify and separate.  From  Fig. 4(e) and (f), it can be seen that ETIF$_1$ and ETCIFLS  improve the energy concentration highly, and IF trajectories of all components of the mixed signal can be identified. Moreover, the ETCIFLS   has better TF concentration and continuity with less noise interference compared with ETIF$_1$.

\begin{table}[!tbph]
	\vspace{0cm}
	\caption{ \label{tab4} Performance comparison of the six TF representations  for  signal (55). } \centering
	\begin{tabular}{ccccccccccc}
		\Xhline{.6pt} Method&{SET}&{SSST}&{SSEWT}&{MDD}&{ETIF$_1$}&{ETCIFLS }\\
		\hline
		Time (s)&0.030 &0.207&0.043&0.274 &0.176&2.588\\
		RE & 11.485 &  12.257  &  10.291  &  15.232 &  12.690 &  11.068\\	
		\Xhline{.6pt}  
	\end{tabular}
	\setlength{\belowcaptionskip}{-0.15cm}
\end{table}
Table \uppercase\expandafter{\romannumeral2}  provides the quantitative results of the employed methods. It can be observed that the first five methods are more efficient in computation. The  ETCIFLS   gives an accurate TF  description and has better TF concentration than SET, SSST, MDD, and ETIF$_1$.     

\subsection{IF detection under different noise  levels}
In order to explore the performance of the proposed approaches in IF characterization and noise robustness, we give the  IF  estimates of a multi-component signal under different  SNRs of noise. The IF trajectories of the signal are detected by linking the peak data from the TF  representation, as used in  [44,49]. The test signal model is defined as:
\begin{equation}
	\begin{split}
		f(t)=& f_1(t)+f_2(t)+n(t),  ~  t\in[0, 3],\\
		f_1(t)=& 0.5\exp(-0.01t)\cos(2\pi (225t-7(t-0.8)^3+5t^2)),\\
		f_2(t)=&\sin(2\pi(90t + 14\arctan((5t^2 - 2)^2))),  
	\end{split}
\end{equation}
where $n(t)$ denotes the Gaussian white noise. The sampling frequency is  512 Hz.

Fig. 5 shows the logarithmic mean square error (MSE) between the true and estimated IF  under different SNRs by performing 100 simulations. The SSEWT fails to produce correct IF approximates of the test signal (56). The MDD shows good noise robustness at SNRs less than 0 dB but fails to improve the accuracy at higher SNRs. The SET and SSST give better representation for $f_1(t)$  than  $f_2(t)$, and the SSST has better performance than SET in dealing with $f_1(t)$. The  ETIF$_1$ is comparable to the performance of SST and SSST at low SNRs (SNR $<$ 0 dB) and generally,  shows better noise robustness than ETCIFLS.  Both ETIF$_1$ and ETCIFLS achieve better performance than other methods when addressing the mode $f_2(t)$ at SNRs greater than 2 dB.  The comparison result indicates the effectiveness of the proposed methods in characterizing signals (56), especially under high SNR. The noise robustness of  ETCIFLS and more effective IF detection algorithms should be explored in later research.      
\begin{figure}[!tbh]
	\centering
	\begin{minipage}{0.49\linewidth}
		\centerline{\includegraphics[width=1\textwidth]{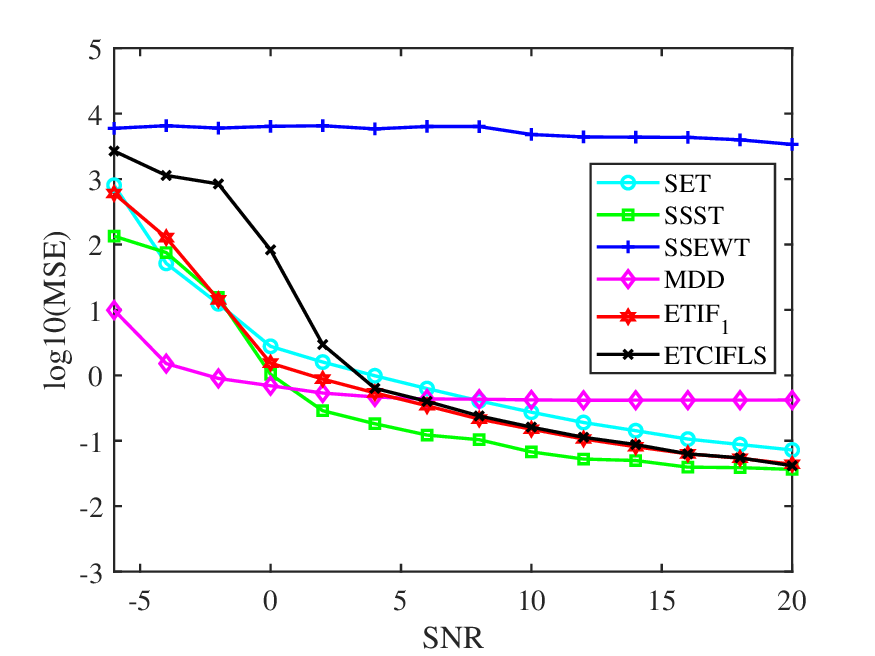}}
		\centerline{(a)}
	\end{minipage}
	\begin{minipage}{0.49\linewidth}
	\centerline{\includegraphics[width=1\textwidth]{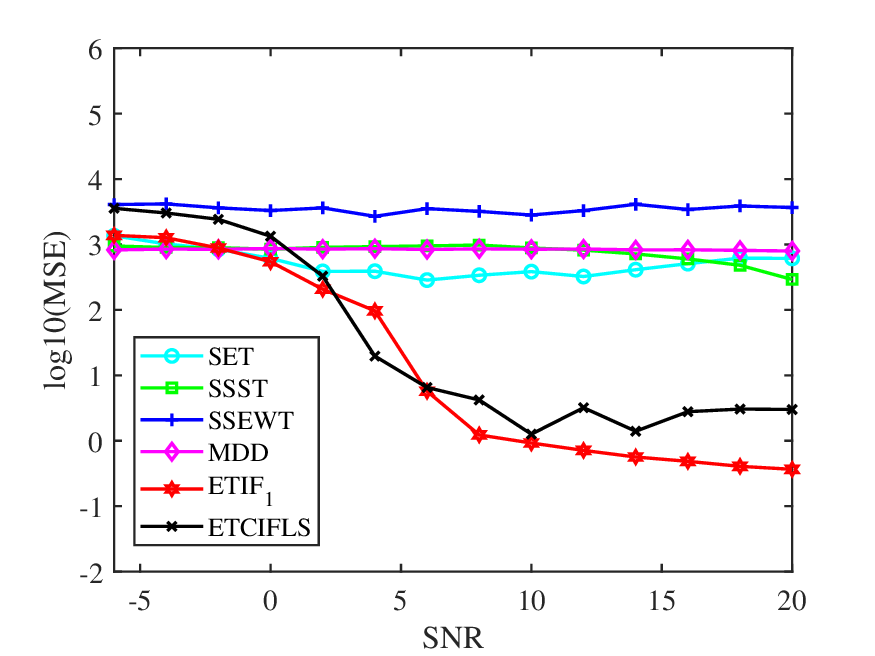}}
	\centerline{(b)}
\end{minipage}
\caption{ \small MSE of the IF estimates of  signal (56) under  different noise interference. (a) MSE  of the  IF estimate of $f_1(t)$, (b)  MSE  of the  IF estimate of $f_2(t)$.}
\end{figure}


\subsection{Application to EEG Seizure Signal}
In this subsection, we will investigate the applicability of the proposed approach for analyzing an epileptic EEG seizure signal [60]. This signal mainly includes  two types of components,  tones and  spikes.  The sampling frequency of the signal is  32 Hz and about 8 seconds of data is displayed in  Fig. 6.   
\begin{figure}[!bth]
	\centering
	\vspace{-0.2cm}
	\begin{minipage}{1\linewidth}
		\centerline{\includegraphics[width=1\textwidth]{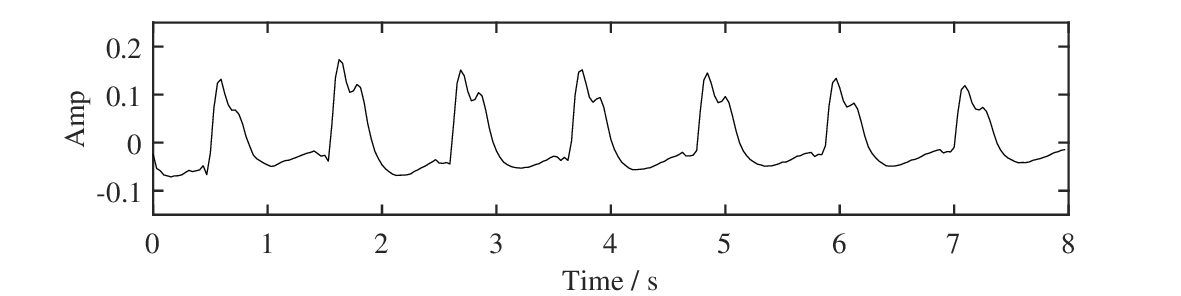}}
	\end{minipage}
	\caption{ \small The waveform of the epileptic EEG seizure signal.}
\end{figure}
\begin{figure}[bth!]
	\vspace{-0.3cm}
	\setlength{\belowcaptionskip}{-0.1cm}
	\centering
	\begin{minipage}{0.49\linewidth}
		\centerline{\includegraphics[width=1\textwidth]{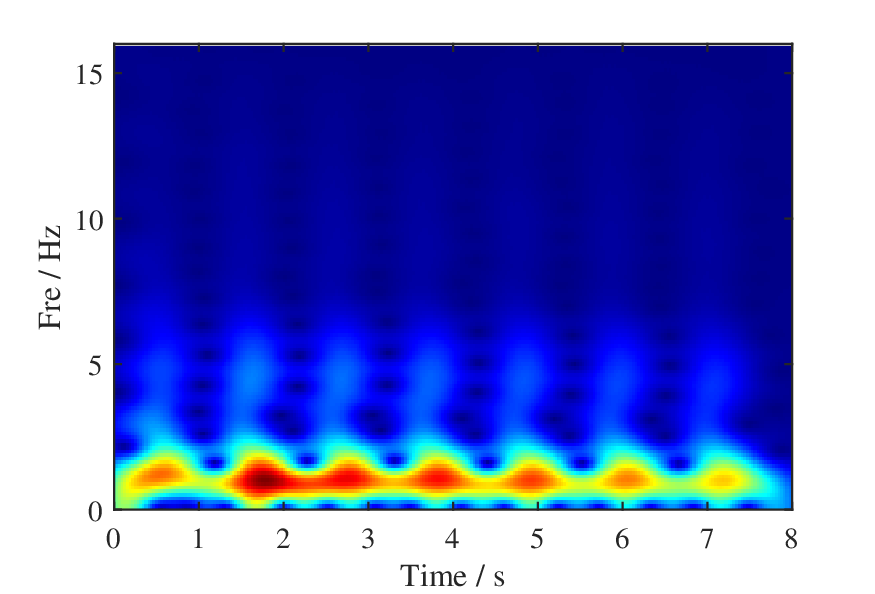}}
		\centerline{(a)}
	\end{minipage}
	\begin{minipage}{0.49\linewidth}
		\centerline{\includegraphics[width=1\textwidth]{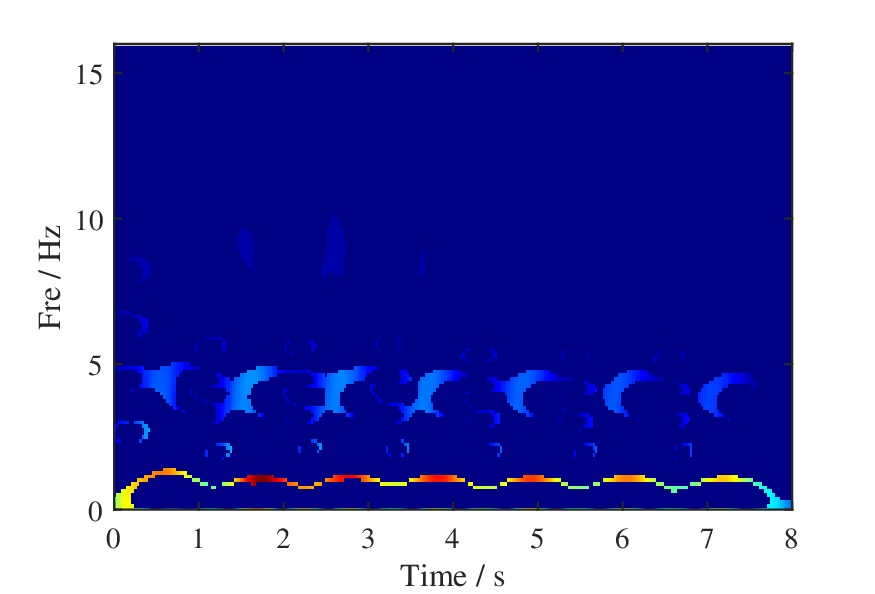}}
		\centerline{(b)}
	\end{minipage} \begin{minipage}{0.49\linewidth}
		\centerline{\includegraphics[width=1\textwidth]{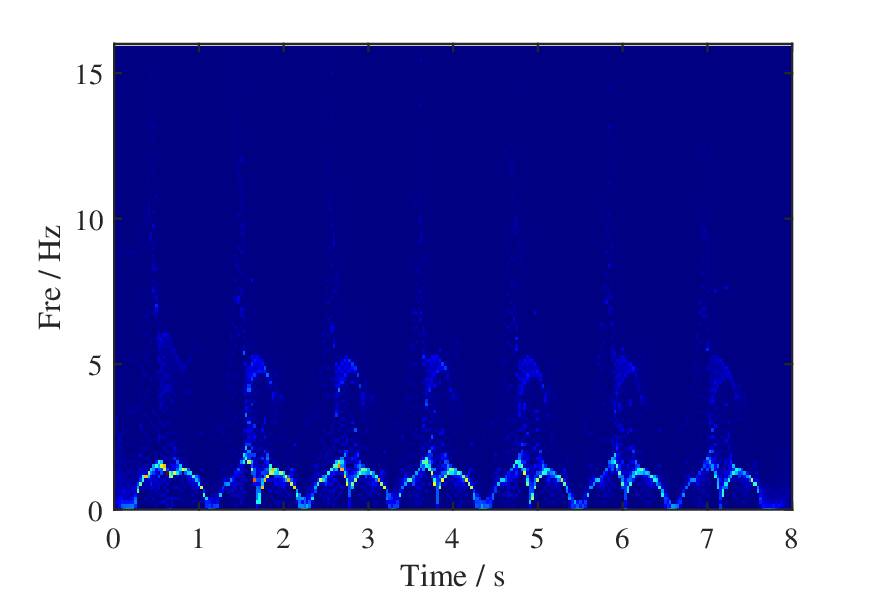}}
		\centerline{(c)}
	\end{minipage}
	\begin{minipage}{0.49\linewidth}
		\centerline{\includegraphics[width=1\textwidth]{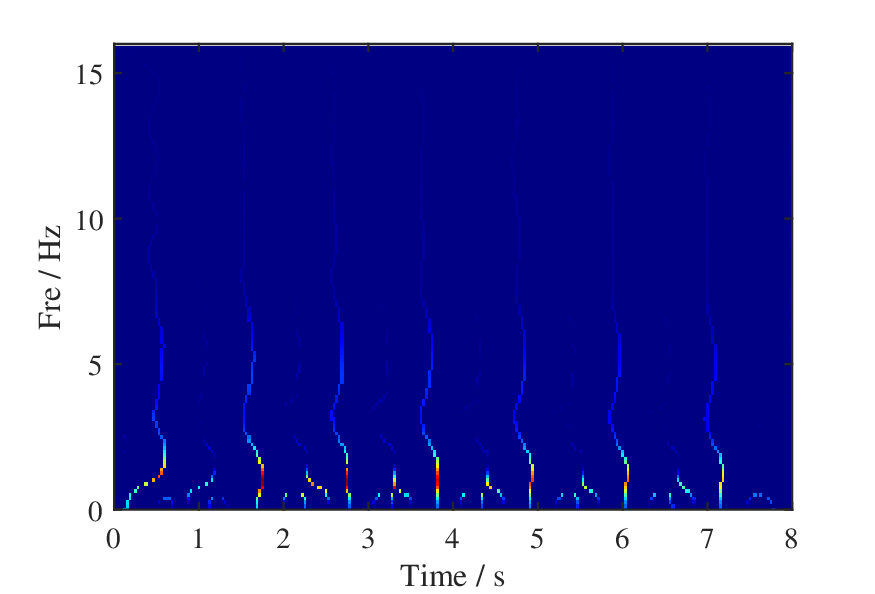}}
		\centerline{(d)}
	\end{minipage}
	\begin{minipage}{0.49\linewidth}
		\centerline{\includegraphics[width=1\textwidth]{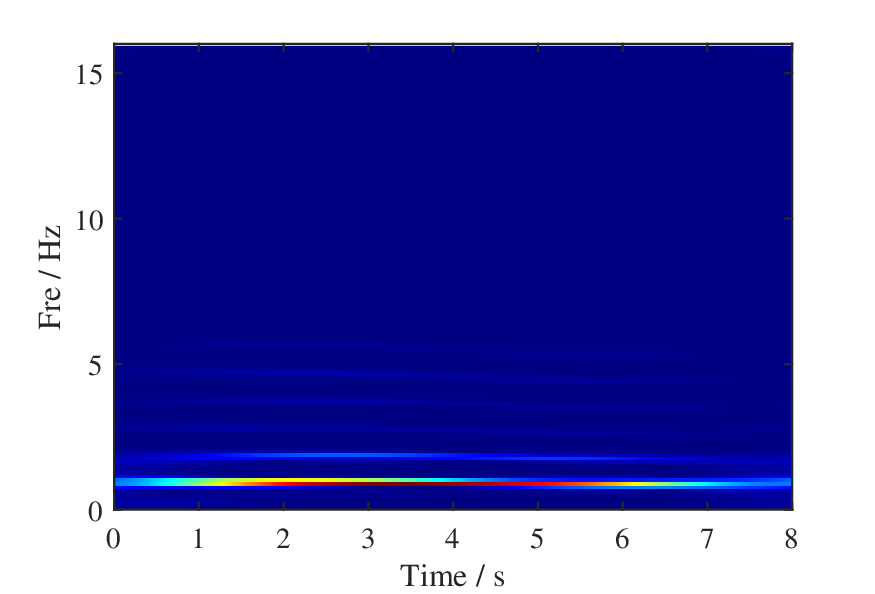}}
		\centerline{(e)}
	\end{minipage}
	\begin{minipage}{0.49\linewidth}
		\centerline{\includegraphics[width=1\textwidth]{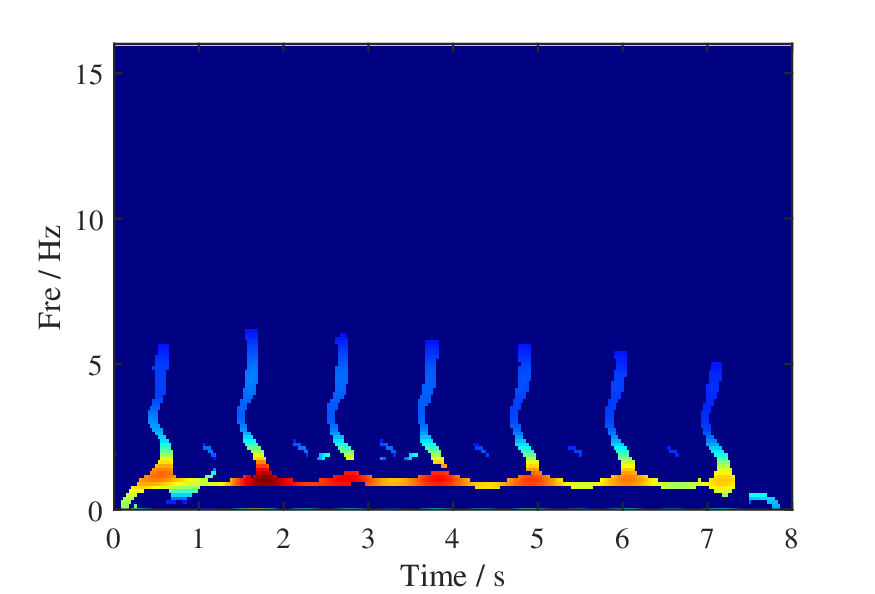}}
		\centerline{(f)}
	\end{minipage}
	\caption{\small TF representations of a real newborn EEG seizure signal composed of  spikes and a pseudosinusoid.   (a)  TF plot by STFT, (b) TF plot by SET,  (c) TF plot by SSST,  (d) TF plot by SSEWT,  (e)  TF plot by  MDD (2, 0.05), (f) TF plot by   ETCIFLS  ($n_w=71$). }
\end{figure}

Fig. 7 shows the corresponding TF representations associated with the STFT, SET, SSST, SSWET, MDD, and  ETCIFLS.  It can be seen that the STFT provides energy characterization for tones, giving blurred TF representation for spikes.  Fig. 7(b)  and (e) show that  SET and MDD sharpen the TF distribution of the tones compared with the STFT, but still fail to resolve the spikes. The SSST (Fig. 7(c)) presents a blurry TF representation, based on which, it is difficult to make further analysis and judgment.  The SSEWT  (Fig. 7(d)) concentrates the TF energy of the spikes while failing to show the TF information of the tones. Fig. 7(f) shows that the proposed  ETCIFLS  can identify both tones and spikes clearly in the TF plane,  exhibiting great potential in EEG  signal analysis.

\subsection{Application to Gravitational-wave Signal}

The second practical signal is a gravitational-wave (GW) signal [61] generated by the merge of two stellar-mass black holes. This  signal is a typical AM-FM signal and  accompanied by  large noise.  The sampling frequency is 4096 and the  time-series waveform of the signal is plotted  in Fig. 8.   
\begin{figure}[!tbh]
	\centering
	\begin{minipage}{1\linewidth}
		\centerline{\includegraphics[width=1\textwidth]{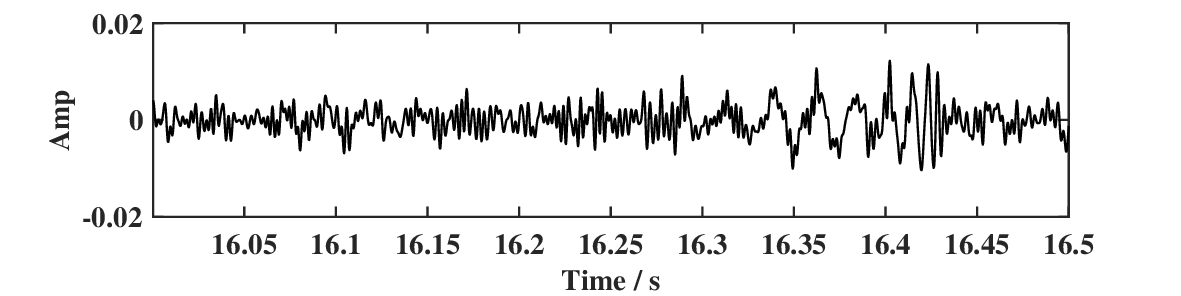}}
	\end{minipage}
	\caption{ \small Waveform of the gravitational-wave signal.}
\end{figure}


The instantaneous information of the GW signal contains significant features corresponding to the collision of two black holes, thus it is necessary to remove any additional spurious disturbance to obtain its accurately time-varying characteristics.  Fig. 9 shows the TF representations of the GW signal by the six TF analysis methods. From the TF plot, we can see it 'chirping' from lower to higher frequency over a small fraction of a second, i.e.,  the hyperbolic-like frequency-modulation component indicated by a yellow box.  For this component, the first three methods yield a spread or absent  TF content for the fast-varying part, while SSEWT provides a  discontinuous TF curve for the harmonic-like part. The MDD  can characterize the basic features of the GW signal. The ETCIFLS shows a good resolution on the TF representation of the hyperbolic-like component and effectively indicates the time of the incoming GW signal. 

The IF estimates of the component at the area of time $[16.3 ~ 16.45]$ are also given in the right half of Fig. 9. It can be seen from the results that the ETCIFLS provides a better characterization, based on which the IF  information can be clearly identified. This case demonstrates the capability of ETCIFLS  to detect the complex gravitational-wave signal and to generate high-concentration TF description in dealing with hyperbolic-like signals.   
\begin{figure}[tbh!]
	\vspace{-0.2cm}
	\setlength{\belowcaptionskip}{-0.1cm}
	\centering
	\begin{minipage}{0.49\linewidth}
		\centerline{\includegraphics[width=1\textwidth]{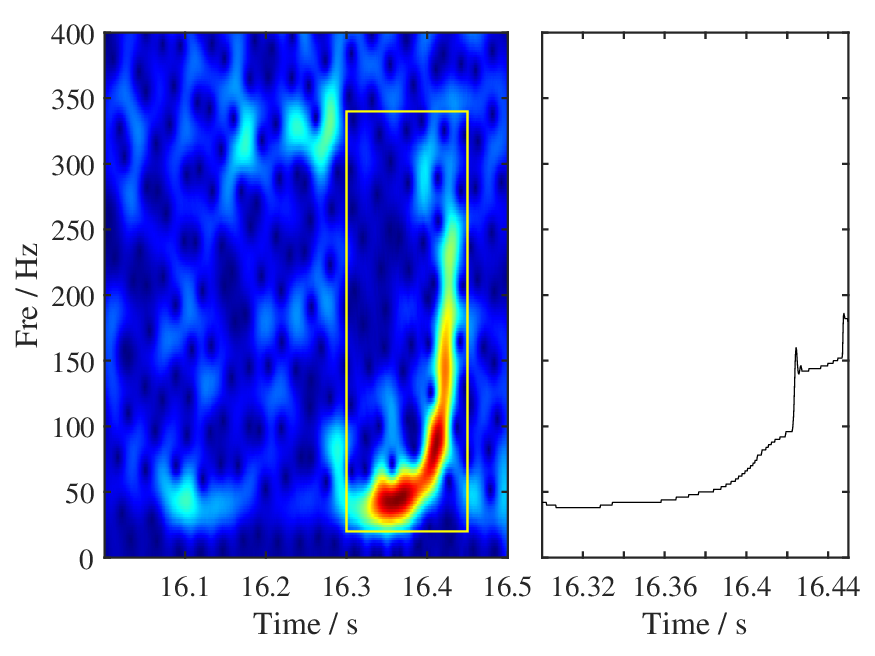}}
		\centerline{(a)}
	\end{minipage}
	\begin{minipage}{0.49\linewidth}
		\centerline{\includegraphics[width=1\textwidth]{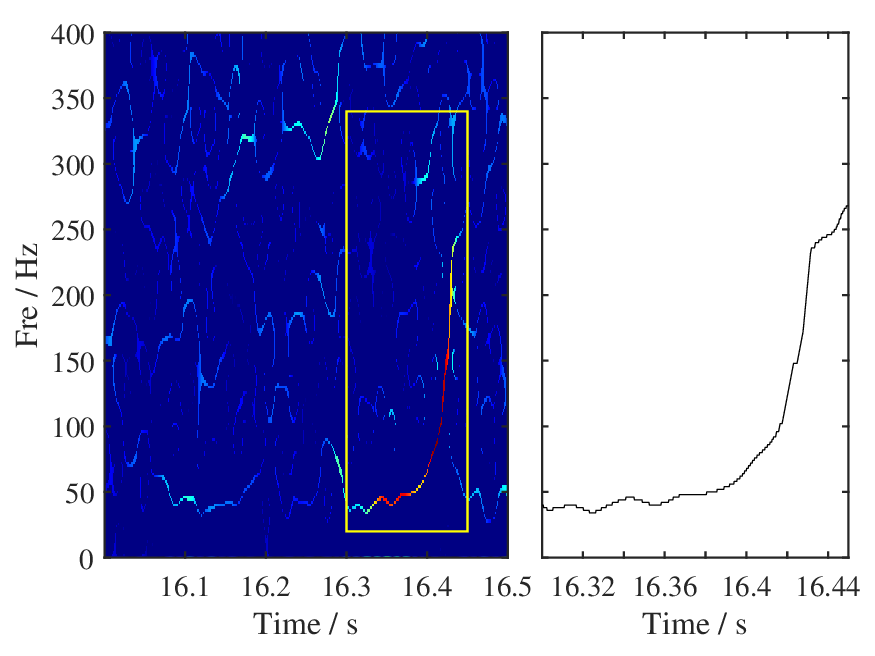}}
		\centerline{(b)}
	\end{minipage} \begin{minipage}{0.49\linewidth}
		\centerline{\includegraphics[width=1\textwidth]{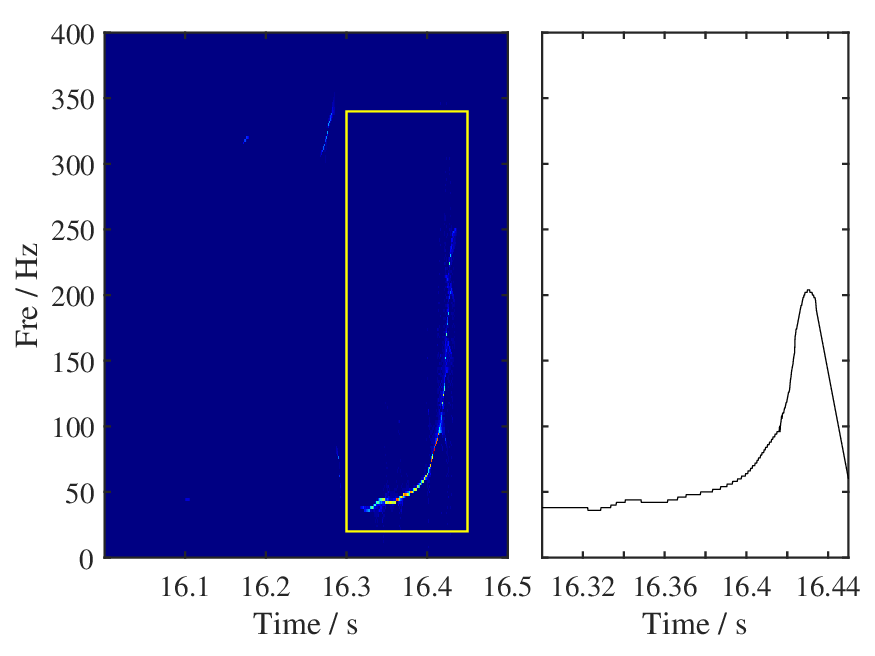}}
		\centerline{(c)}
	\end{minipage}
	\begin{minipage}{0.49\linewidth}
		\centerline{\includegraphics[width=1\textwidth]{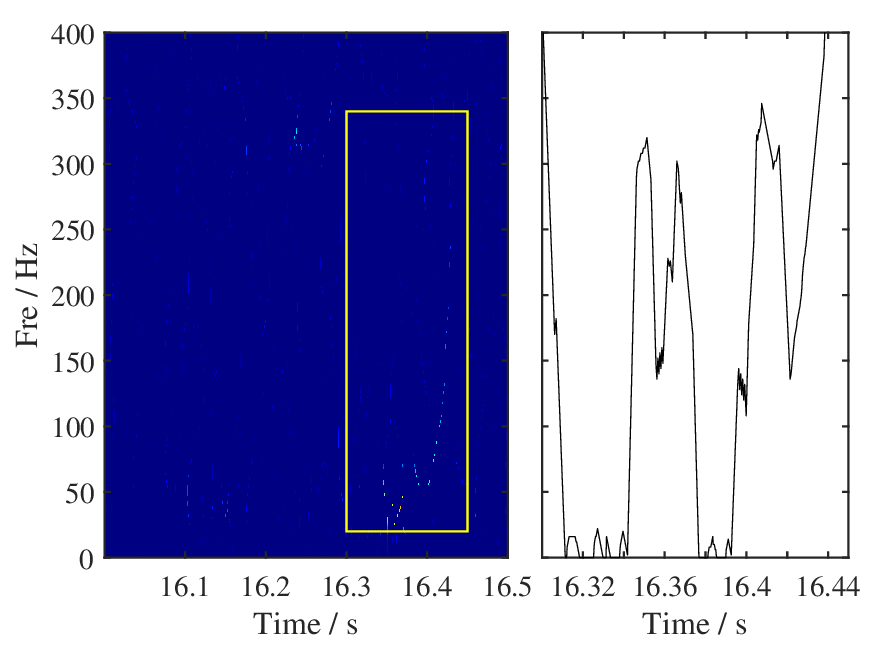}}
		\centerline{(d)}
	\end{minipage}
	\begin{minipage}{0.49\linewidth}
		\centerline{\includegraphics[width=1\textwidth]{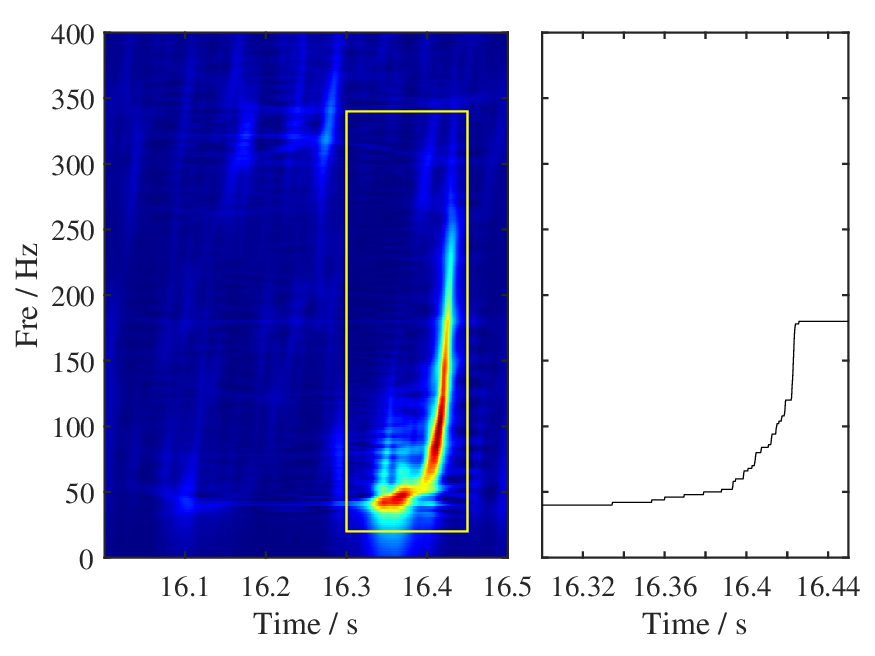}}
		\centerline{(f)}
	\end{minipage}
	\begin{minipage}{0.49\linewidth}
		\centerline{\includegraphics[width=1\textwidth]{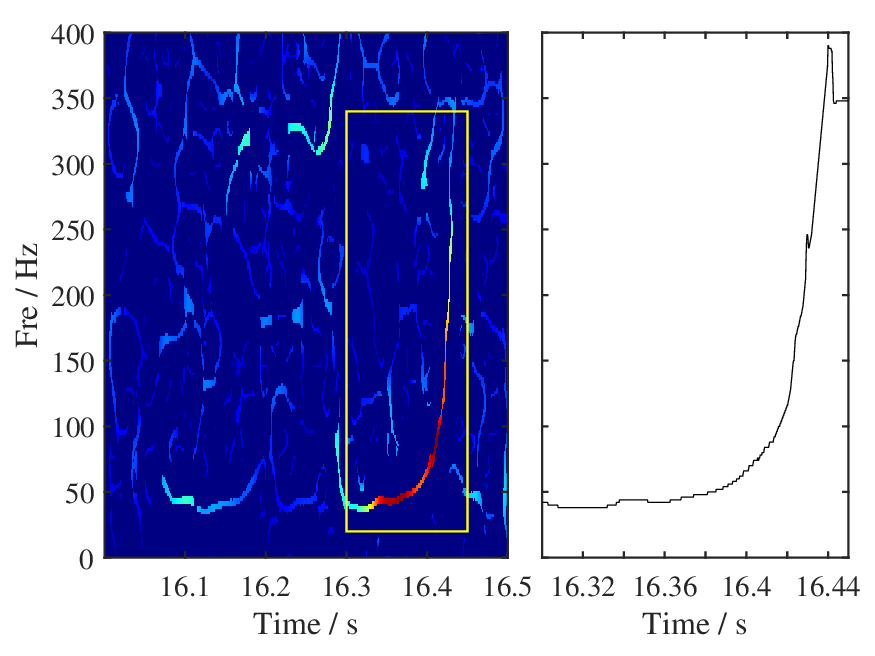}}
		\centerline{(g)}
	\end{minipage}
	\caption{ \small   TF results and IF estimation  by various analysis methods.  (a)   STFT, (b) SET,  (c) SSST,  (d)  SSEWT,  (e)    MDD (2, 0.05), (f) ETCIFLS ($n_w=51$). }
\end{figure}

\section{Conclusion}
In this paper, we discussed the IF equation-based TF analysis method to generate a concentrated TF representation for mixed signals.   The IF equation we defined does serve as a feature extractor to characterize the slow-varying and fast-varying modes. Several properties of the IF equation were studied and proved theoretically. Based on the IF equation, we introduced two classical ways, that is,  extraction and reassignment, to obtain a highly-concentrated TF output. The IF equation-based TF post-processing methods provide another way to improve the TF resolution and cover many popular TF analysis methods, such as the synchroextracting transform, the multi-synchrosqueezing transform, and the time extracting transform, etc.  We also proposed combining different IF equations based on local information to preserve the best concentration of each representation. Several experiments on both synthetic and real-world signals showed that our approaches have better performance in dealing with a  mixture of impulse-like signals, hyperbolic frequency modulation components,  chirps, and cosine modulation signals. 

We are going to share the corresponding code on the Matlab Central website soon, and the code is also available upon the
request to the first author.


\end{document}